\documentclass[letterpaper]{article} 
\usepackage{aaai24}  
\usepackage{times}  
\usepackage{helvet}  
\usepackage{courier}  
\usepackage[hyphens]{url}  
\usepackage{graphicx} 
\usepackage{natbib}  
\usepackage{caption} 
\usepackage{algorithm}
\usepackage{algorithmic}

\usepackage{amsmath}
\usepackage{amssymb}
\usepackage{hyperref}
\hypersetup{
    colorlinks=true,
    breaklinks=true,
    citecolor={green!70!black},
}
\usepackage[capitalize,noabbrev]{cleveref}
\usepackage{multirow}
\usepackage{color}
\usepackage{subfigure}
\usepackage{threeparttable}
\usepackage{array}
\usepackage{booktabs}
\usepackage{enumitem}
\usepackage{xcolor}

\makeatletter
\def\@fnsymbol#1{\ensuremath{\ifcase#1\or \dagger\or \ddagger\or
   \mathsection\or \mathparagraph\or \|\or **\or \dagger\dagger
   \or \ddagger\ddagger \else\@ctrerr\fi}}
\makeatother

\usepackage{newfloat}
\usepackage{listings}
\DeclareCaptionStyle{ruled}{labelfont=normalfont,labelsep=colon,strut=off} 
\floatstyle{ruled}
\newfloat{listing}{tb}{lst}{}
\floatname{listing}{Listing}
\title{Settling Decentralized Multi-Agent Coordinated Exploration by Novelty Sharing}

\author {
    Haobin Jiang\textsuperscript{\rm 1},
    Ziluo Ding\textsuperscript{\rm 1,\rm 2},
    Zongqing Lu\textsuperscript{\rm 1}\footnote{Corresponding author}
}
\affiliations {
    \textsuperscript{\rm 1} School of Computer Science, Peking University \\
    \textsuperscript{\rm 2} Beijing Academy of Artificial Intelligence \\
    \texttt{\{haobin.jiang, ziluo, zongqing.lu\}@pku.edu.cn}
}

\usepackage{bibentry}

\begin{document}

\maketitle

\begin{abstract}
    Exploration in decentralized cooperative multi-agent reinforcement learning faces two challenges. One is that the novelty of global states is unavailable, while the novelty of local observations is biased. The other is how agents can explore in a coordinated way. To address these challenges, we propose MACE, a simple yet effective multi-agent coordinated exploration method. By communicating only local novelty, agents can take into account other agents' local novelty to approximate the global novelty. Further, we newly introduce weighted mutual information to measure the influence of one agent's action on other agents' accumulated novelty. We convert it as an intrinsic reward \textit{in hindsight} to encourage agents to exert more influence on other agents' exploration and boost coordinated exploration. Empirically, we show that MACE achieves superior performance in three multi-agent environments with sparse rewards. The code is available at \href{https://github.com/SigmaBM/MACE}{https://github.com/SigmaBM/MACE}.
\end{abstract}

\section{Introduction}
\label{intro}

Recent progress in decentralized learning theories and algorithms for multi-agent reinforcement learning (MARL) \citep{zhang2018fully,de2020independent,Jin2021VLearningA,Daskalakis2022TheCO,jiang2022i2q} makes it feasible to learn high-performant policies in a decentralized way for \textit{cooperative} multi-agent tasks. However, one critical issue remains, \textit{i.e.}, how to enable agents to effectively explore in a coordinated way under such a learning paradigm, especially for sparse-reward tasks where the environment rarely provides rewards.  

One of the most popular exploration schemes in the single-agent setting is novelty-based exploration \cite{bellemare2016unifying,pathak2017curiosity,burda2018exploration,zhang2021noveld}, where the agent is encouraged by well-designed intrinsic reward to visit \textit{novel} states it rarely sees. However, things could be different when migrating to decentralized multi-agent settings, which leads to an unsolved problem: as only the local observation instead of the global state is available, how should each agent measure the novelty of the global state?  

In decentralized settings, partial observability expands the discrepancy between each agent's local observation novelty and global state novelty, which makes the exploration merely based on local novelty highly unreliable \cite{wang2019influence,iqbal2019coordinated}. Fortunately, communication can help ease partial observability by providing extra information about other agents \cite{jiang2018learning,das2019tarmac,wang2019learning,ding2020learning}. However, unlimited communication may incur too much communication overhead, and it can indeed transform the decentralized setting into a centralized setting. Therefore, we resort to decentralized learning with \textit{limited communication} to address such problems.

In addition to the challenge of novelty measurement, in cooperative tasks, agents must also acquire the ability to coordinate with each other to explore and achieve the final goal.
Ideally, the optimal exploration strategy should consider others' observations and actions. Previous work \cite{wang2019influence,iqbal2019coordinated,liu2021cooperative} finds that independent exploration is not efficient and redundant exploration occurs. By coordination in exploration, we mean agents help other agents to achieve novel observations or reach novel states together through cooperation. In other words, an agent should be encouraged when its action enables other agents to reach more novel observations.

In this paper, we propose a \textit{simple yet effective} \textbf{M}ulti-\textbf{A}gent \textbf{C}oordinated \textbf{E}xploration method, namely \textbf{MACE}. MACE introduces a \textit{novelty-based} intrinsic reward and a \textit{hindsight-based} intrinsic reward to enable coordinated exploration in decentralized cooperative tasks. Within the confines of limited communication, agents only share their local novelties (merely a floating point number) during training. Each agent leverages this shared information to approximate the global novelty, which serves as the novelty-based intrinsic reward. This approach aims to bridge the gap between the local novelty and the global novelty.
Moreover, we encourage agents to exert more influence on others’ explorations through the hindsight-based intrinsic reward, thereby boosting coordinated exploration. To this end, we measure the weighted mutual information \cite{Guiasu1977-rs,schaffernicht2011weighted} between the action of the agent and the accumulated novelty obtained thereafter by others given the local observation. The higher the weighted mutual information value, the higher the hindsight-based intrinsic reward the agent receives.

We evaluate MACE in three multi-agent environments: GridWorld, Overcooked \cite{carroll2019utility}, and SMAC \cite{samvelyan2019starcraft}. All tasks in these environments are sparse-reward and hard to explore. The experimental results verify the effectiveness of MACE. Through ablation studies, we show that both the approximation to global novelty and the encouragement to influence other agents' exploration are indispensable in decentralized multi-agent exploration, and our newly employed weighted mutual information works significantly better than normal mutual information.

\section{Preliminary}
\label{prel}

\textbf{Decentralized learning.} We consider an \textit{N}-agent Markov decision process (MDP) $\mathcal{M}=\{S,\boldsymbol{O},\boldsymbol{A},P,R,\gamma\}$. Here, $S$ represents the state space while $\boldsymbol{A}$ is the joint action space; the transition probability is defined by $P(s'|s,\boldsymbol{a})$. In decentralized learning, each agent has only access to its own local observation $o_i\in O_i$, rather than the global state, and learns an independent policy $\pi_i$ to maximize the shared reward defined by $R$ together with other agents. Notably, decentralized learning is more practical than centralized learning, owing to its better scalability, privacy, and security \cite{Zhang2019multi}.

\vspace{2mm}
\noindent \textbf{Limited communication.} We allow agents to communicate during the training phase. However, in order to enhance the practicality and adhere closely to the decentralized setting, we impose constraints on the bandwidth of the communication channel to reduce communication overhead \cite{foerster2016learning,kim2018learning,wang2020learning}. Specifically, the message sent by one agent at each step is confined to a floating point number. In this paper, we set agents to communicate their local novelties through this limited channel. Communication is not allowed during execution. 

Note that this setting differs from centralized training and decentralized execution (CTDE) \cite{lowe2017multi,COMA,rashid2018qmix}, where agents can use unlimited extra information to ease training, such as other agents' observations and actions, or a centralized value function. Besides, our setting is not identical to \textit{fully} decentralized learning \cite{tan1997MultiAgentRL,de2020independent,jiang2022i2q}, where communication is forbidden. On top of the fully decentralized learning algorithm, we will show that adding communication of novelty during training can enable coordinated exploration of agents to solve sparse-reward tasks.

\section{Methodology}

In this section, we present MACE addressing the challenges in decentralized multi-agent exploration. MACE follows the line of intrinsically motivated exploration \cite{yang2021exploration} that designs intrinsic rewards $r_\mathrm{int}$ and trains agents via the shaped reward $r_s=r_\mathrm{ext}+r_\mathrm{int}$, where $r_\mathrm{ext}$ denotes the extrinsic reward given by the environment. MACE adopts the following two parts to design intrinsic rewards: 1) To obtain a more reliable novelty estimate as the novelty-based intrinsic reward, MACE uses the summation of all agents' novelty to approximate the global novelty (\cref{sec:3_1}). 2) To boost coordinated exploration, MACE further quantifies the influence of agents on the accumulated future novelty of other agents (\cref{sec:3_2}) and converts it into the hindsight-based intrinsic reward (\cref{sec:3_3}). The weighted sum of these two parts is the final intrinsic reward used in MACE (\cref{sec:3_4}). 

\subsection{Approximation to Global Novelty}
\label{sec:3_1}

In decentralized training, if we only take into account the individual exploration of agent $i$, $u_t^i=\mathrm{novelty}(o_{t+1}^i)$ can serve as the intrinsic reward to encourage agent $i$ to take actions towards observations it seldom visits. When the observation space is discrete and small, such as the 2-dimension grid $(x,y)$, we could directly record the number of times each observation that agent $i$ has visited before and define $\mathrm{novelty}(o)=1/n(o)$ where $n(o)$ denotes the visit counts. If the observation space becomes large or continuous, methods designed for high-dimensional input such as pseudo-count \cite{bellemare2016unifying}, ICM \cite{pathak2017curiosity}, and RND \cite{burda2018exploration} could be used to measure the novelty. 

However, $u_t^i$ only measures the local novelty of agent $i$. In the multi-agent environment, given the discrepancy between the local novelty and the global novelty, $u_t^i$ may not be able to provide accurate and sufficient information for exploration. For example, we consider a two-agent environment where at timestep $t$, agent 1 is in an observation with low local novelty, and agent 2 is in an observation with high local novelty. From the global perspective, the two agents are in a novel state. However, from agent 1's perspective, it thinks that the observation is not novel and gives itself a low intrinsic reward, preventing it from further exploring the current novel state.

Due to the decentralized setting, the global novelty is not available to each agent. Therefore, we need a more appropriate intrinsic reward term than $u_t^i$ to narrow the gap with the global novelty. Thanks to the limited communication, agents can exchange their local novelty $u_t^i$ with each other at each timestep $t$. We propose a heuristic that uses the summation of all agents' local novelty as an approximation to the global novelty and as the novelty-based intrinsic reward:
\begin{equation}
\label{equ:rn}
    r_\mathrm{nov}^i(o_t^i,a_t^i)=\sum_ju_t^j.
\vspace{-1mm}
\end{equation}
With the introduction of other agents' novelty, we can avoid the aforementioned dilemma. 

We admit that \eqref{equ:rn} still deviates from the global novelty in some cases. For example, agent 1 and agent 2 are in low-novelty observations while the global state is novel, which occurs when agent 1 and agent 2 seldom visit current observations \textit{simultaneously}. Nevertheless, the gap with the global novelty cannot be closed entirely due to the limited information, and experimental results prove empirically that \eqref{equ:rn} works better than the local novelty $u_t^i$ (\cref{sec:experiment}). One may argue that an alternative is to use the maximum of all agents' novelty as the intrinsic reward, but our empirical result shows that \eqref{equ:rn} works better (\cref{app:sum_vs_max}).

\subsection{Influence on Other Agents’ Exploration}

To boost coordinated exploration in multi-agent environments, each agent should consider its influence on other agents' exploration so that it could find some \textit{critical states} \cite{yang2021exploration}. Critical states here mean that in these states, the action taken by one agent affects other agents' exploration progress, \textit{e.g.}, one agent steps on a switch and thus opens a door that blocks another agent's way. So encouraging agents to explore these critical states would help them learn to cooperate effectively. We first discuss how to quantify one agent's influence on other agents' exploration.

Suppose there are two agents, agent $i$ and agent $j$, in the environment. To estimate agent $i$'s influence on agent $j$'s exploration in a specific observation, we could use \textit{mutual information}, a common measure used in MARL \cite{li2022pmic}, to quantify the dependence between agent $i$'s action $a^i_t$ and agent $j$'s accumulated novelty $z^j_t=\sum_{t'=t}\gamma^{t'-t}u^j_{t'}$ given agent $i$'s observation $o_t^i$: 
\begin{equation*}
     I\left(A_t^i;Z^j_t | o_t^i\right)=\mathbb{E}_{a_t^i,z_t^j | o_t^i}\left[\log\frac{p(a_t^i,z_t^j | o_t^i)}{p(a_t^i | o_t^i)p(z_t^j | o_t^i)}\right].
\end{equation*}
Here we use agent $j$'s accumulated novelty $z^j_t$ instead of its immediate novelty $u_t^j$ to measure the long-term dependence.

\label{sec:3_2}

\begin{table}[t]
    \caption{Actions and reward probabilities of two illustrative states.}
    \label{tab:illu}
    
    \centering
    \renewcommand{\arraystretch}{1.2}
    \begin{footnotesize}
    \begin{tabular}{|c|c|}
    \multicolumn{2}{c}{\textbf{state 1}} \\
    \multicolumn{1}{c}{act} & \multicolumn{1}{c}{reward probability} \\
    \cline{1-2}
    \multirow{3}*{$a_1$} & $p(r=1\mid a_1)=0.1$ \\
    \cline{2-2}
     & \textcolor{red}{$p(r=5\mid a_1)=0.8$} \\
    \cline{2-2}
     & $p(r=9\mid a_1)=0.1$ \\
    \cline{1-2}
    \multirow{3}*{$a_2$} & \textcolor{red}{$p(r=1\mid a_2)=0.8$} \\
    \cline{2-2}
     & $p(r=5\mid a_2)=0.1$ \\
    \cline{2-2}
     & $p(r=9\mid a_2)=0.1$ \\
    \cline{1-2}
    \end{tabular}
    \hspace{0.1in}
    \begin{tabular}{|c|c|}
    \multicolumn{2}{c}{\textbf{state 2}} \\
    \multicolumn{1}{c}{act} & \multicolumn{1}{c}{reward probability} \\
    \cline{1-2}
    \multirow{3}*{$a_1$} & $p(r=1\mid a_1)=0.1$ \\
    \cline{2-2}
     & \textcolor{red}{$p(r=5\mid a_1)=0.8$} \\
    \cline{2-2}
     & $p(r=9\mid a_1)=0.1$ \\
    \cline{1-2}
    \multirow{3}*{$a_2$} & $p(r=1\mid a_2)=0.1$ \\
    \cline{2-2}
     & $p(r=5\mid a_2)=0.1$ \\
    \cline{2-2}
     & \textcolor{red}{$p(r=9\mid a_2)=0.8$} \\
    \cline{1-2}
    \end{tabular}
    \end{footnotesize}
\end{table}

\begin{figure}[t]
    \centering
    \centering
    \subfigure[MI]{
        \includegraphics[width=0.45\columnwidth]{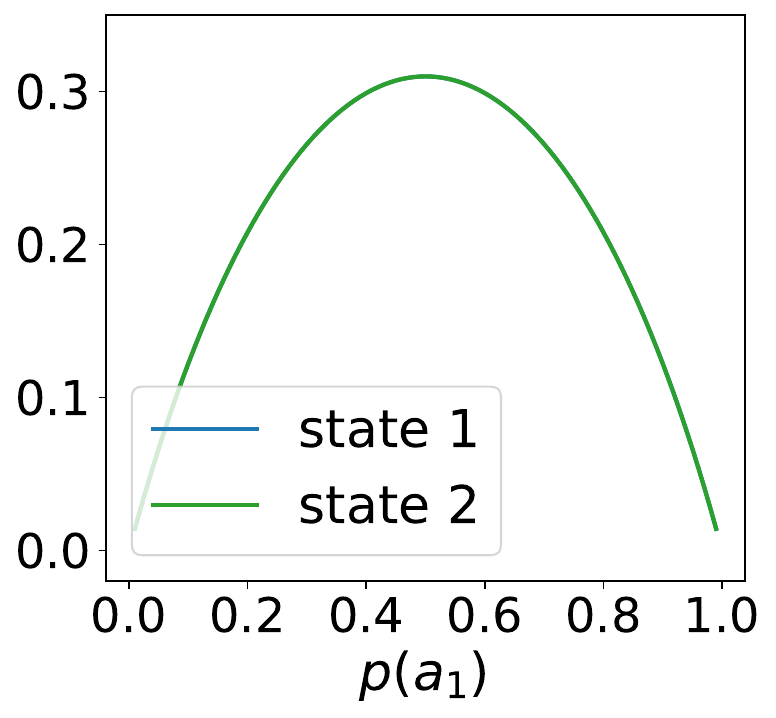}
        \label{fig:illu_mi}
    }
    \hfill
    \subfigure[WMI]{
        \includegraphics[width=0.45\columnwidth]{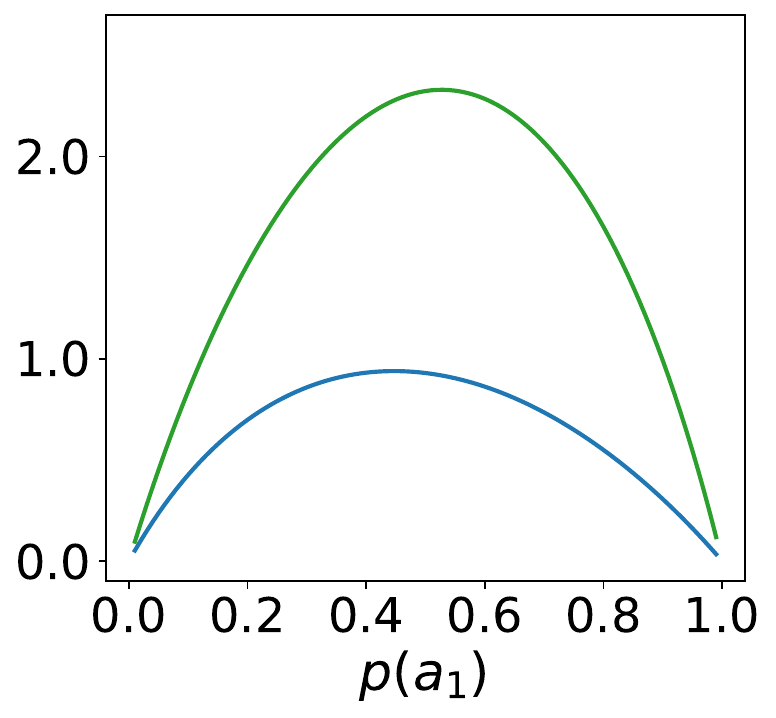}
        \label{fig:illu_wmi}
    }
    \caption{(a) Mutual information (MI) between action and reward in state 1 and state 2. (b) Weighted mutual information (WMI) between action and reward in state 1 and state 2.}
    \label{fig:illu}
\end{figure}

However, mutual information ignores the \textit{magnitude} of agent $j$'s accumulated novelty $z_t^j$. So it would give similar measurements for observation $o_1^i$ where $a_t^i$ is related to some low-value $z_t^j$, and observation $o_2^i$ where $a_t^i$ is related to some high-value $z_t^j$. To illustrate the statement intuitively, we devise two states with two actions and three different rewards, described in \cref{tab:illu}. Action and reward here can be seen as $a_t^i$ and $z_t^j$ respectively. As shown in \cref{fig:illu_mi}, with different $p(a_1)$, state 1 and state 2 always keep the same mutual information. But state 2 is more critical to coordinated exploration because the action ($a_2$) taken by agent $i$ in state 2 can lead agent $j$ to high accumulated novelty ($r = 9$) more likely. Although agent $i$'s action in state 1 has an influence on agent $j$'s accumulated novelty to the same extent as that in state 2 measured by mutual information, it is more likely to result in lower accumulated novelty ($r = 1$ or $r = 5$).  Therefore we need a more effective measure to estimate the influence of agent $i$'s action $a_t^i$ on agent $j$'s accumulated novelty $z_t^j$, while taking into account the magnitude of $z_t^j$.

To this end, we newly introduce \textit{weighted mutual information} \cite{Guiasu1977-rs,schaffernicht2011weighted} between agent $i$'s action $a^i_t$ and agent $j$'s accumulated novelty $z^j_t$ given agent $i$'s observation $o_t^i$: 
\begin{align*}
    \omega I\left(A_t^i;Z^j_t |\right.&\left.\!o_t^i\right) = \\
    & \mathbb{E}_{a_t^i,z_t^j | o_t^i}\left[\omega(a_t^i,z_t^j)\log\frac{p(a_t^i,z_t^j | o_t^i)}{p(a_t^i | o_t^i)p(z_t^j | o_t^i)}\right],
\end{align*}
where $\omega(\cdot,\cdot)$ denotes the weight placed on the pair of $a_t^i$ and $z_t^j$. By introducing weights, pairs of $a_t^i$ and $z_t^j$ would have different informativeness. We set $\omega(a_t^i,z_t^j)=z_t^j$, meaning that relational mappings between $a_t^i$ and higher $z_t^j$ carry more significance than others.
Then, the final measure we use can be written as:
\begin{equation}
\label{equ:wmi}
    \omega I\left(A_t^i;Z^j_t | o_t^i\right) =
    \mathbb{E}_{a_t^i,z_t^j| o_t^i}\left[z_t^j\log\frac{p(a_t^i,z_t^j| o_t^i)}{p(a_t^i| o_t^i)p(z_t^j | o_t^i)}\right].
\end{equation}

To illustrate how it works, \cref{fig:illu_wmi} shows weighted mutual information of state 1 and state 2 with different $p(a_1)$, where we can observe that the weighted mutual information of state 2 is always higher than that of state 1, consistent with what we expected. To summarize, \eqref{equ:wmi} evaluates an observation $o^i_t$ based not only on whether agent $i$'s action has an influence on agent $j$'s exploration, but also on whether agent $i$'s action could lead to a high accumulated novelty of agent $j$.

\subsection{Intrinsic Reward in Hindsight}
\label{sec:3_3}

To encourage each agent $i$ to visit observations with high weighted mutual information, we define an intrinsic reward $r_\mathrm{wmi}^i$ of its observation $o_t^i$ as:
\begin{align}
    r_\mathrm{wmi}^i(o^i_t)
    &=\sum_{j\neq i} r_\mathrm{wmi}^{i\rightarrow j}(o_t^i), \\
    \label{equ:ri_o}
    r_\mathrm{wmi}^{i\rightarrow j}(o_t^i)
    &=\omega I(A_t^i;Z_t^j | o_t^i).
\end{align}
$r_\mathrm{wmi}^{i\rightarrow j}$ denotes the intrinsic reward given to agent $i$ corresponding to its influence on agent $j$'s exploration measured by weighted mutual information. Agent $i$'s intrinsic reward $r_\mathrm{wmi}^i$ is the summation of all $r_\mathrm{wmi}^{i\rightarrow j}$, representing its total influence on other agents' exploration. However, it is nontrivial to compute $r_\mathrm{wmi}^{i\rightarrow j}$ according to \eqref{equ:wmi}, because it is an expectation over all actions and accumulated novelty. So we
decompose the intrinsic reward \eqref{equ:ri_o} onto each action:
\begin{equation}
\label{equ:ri_oa}
    r_\mathrm{wmi}^{i\rightarrow j}(o_t^i,a_t^i)=\mathbb{E}_{z_t^j | o_t^i,a_t^i}\left[z_t^j\log\frac{p(a_t^i,z_t^j | o_t^i)}{p(a_t^i | o_t^i)p(z_t^j | o_t^i)}\right].
\end{equation}
Further, we can continue to decompose \eqref{equ:ri_oa} and obtain a hindsight-based intrinsic reward:
\begin{equation}
\label{equ:ri_oag}
    r_\mathrm{hin}^{i\rightarrow j}(o_t^i,a_t^i,z_t^j)=z_t^j\log\frac{p(a_t^i,z_t^j | o_t^i)}{p(a_t^i | o_t^i)p(z_t^j | o_t^i)}.
\end{equation}
Here $p(a_t^i | o_t^i)$ is the current policy $\pi^i(a_t^i | o_t^i)$ of agent $i$. With Bayesian rule $p(a_t^i | o_t^i, z_t^j)=\frac{p(a_t^i,z_t^j | o_t^i)}{p(z_t^j | o_t^i)}$, we can rewrite the logarithmic term in \eqref{equ:ri_oag} and have:
\begin{align}
\label{equ:final_rij}
r_\mathrm{hin}^{i\rightarrow j}(o_t^i,a_t^i,z_t^j)&=z_t^j\log\frac{p(a_t^i | o_t^i,z_t^j)}{\pi^i(a_t^i | o_t^i)}, \\
\label{equ:final_ri}
r_\mathrm{hin}^i(o_t^i,a_t^i,\{z_t^j\}_{j\neq i})&=\sum_{j\neq i}r_\mathrm{hin}^{i\rightarrow j}(o_t^i,a_t^i,z_t^j),
\end{align}
which are the forms of the hindsight-based intrinsic reward we use in the paper. The term `hindsight' reflects the difference between \eqref{equ:ri_oag} to \eqref{equ:final_ri} and normal reward functions, where the former use information obtained in future, \textit{i.e.}, agent $j$'s accumulated novelty $z_t^j$, which is not available until the end of the episode.

The logarithmic term in \eqref{equ:final_rij} is the \textit{pointwise mutual information} between $a_t^i$ and $z_t^j$. Pointwise mutual information measures the association between two random variables, commonly used in natural language processing (NLP). Therefore, \eqref{equ:final_rij} could be interpreted as encouraging action associative with the high accumulated novelty of agent $j$. If an $a_t^i$ co-occurs with a high $z_t^j$ at timestep $t$ but there is no association between them, the logarithmic term in \eqref{equ:final_rij} will be around zero and agent $i$ will not receive a high intrinsic reward, despite the high $z_t^j$.

Note that the hindsight-based intrinsic rewards keep the following relationship with the original $r_\mathrm{wmi}^{i\rightarrow j}(o_t^i)$:
\begin{align}
\label{equ:mc}
    r_\mathrm{wmi}^{i\rightarrow j}(o_t^i)&=\mathbb{E}_{a_t^i| o_t^i}\left[r_\mathrm{wmi}^{i\rightarrow j}(o_t^i,a_t^i)\right]\nonumber \\ 
    &=\mathbb{E}_{a_t^i,z_t^j | o_t^i}\left[r_\mathrm{hin}^{i\rightarrow j}(o_t^i,a_t^i,z_t^j)\right],
\end{align}
thus using $r_\mathrm{hin}^{i\rightarrow j}(o_t^i,a_t^i,z_t^j)$ could be regarded as a Monte Carlo method for estimating $r_\mathrm{wmi}^{i\rightarrow j}(o_t^i)$. 

To calculate $r_\mathrm{hin}^{i\rightarrow j}(o_t^i,a_t^i,z_t^j)$ at each timestep $t$, agent $i$ needs agent $j$'s accumulated novelty $z_t^j$ and the posterior distribution $p(a_t^i| o_t^i,z_t^j)$. The former is computed at the end of the episode by accumulating agent $j$'s novelty $u_t^j$ that agent $i$ obtained through communication. Note that this does not require additional communication, and each agent still communicates only local novelty at each timestep. The latter could be estimated from trajectory samples. To fulfill \eqref{equ:mc}, samples used to estimate $p(a_t^i| o_t^i,z_t^j)$ and compute  $r_\mathrm{hin}^{i\rightarrow j}(o_t^i,a_t^i,z_t^j)$ should be on-policy because the expectation follows the distribution over $z^j_t$ which is determined by the current policies of all agents. So our proposed intrinsic reward is more suitable for \textit{on-policy} reinforcement learning algorithms.

\subsection{MACE}
\label{sec:3_4}

We combine the novelty-based intrinsic reward \eqref{equ:rn} and the hindsight-based intrinsic reward \eqref{equ:final_ri} to get the final shaped reward:
\begin{align}
\label{equ:final_r}
     r_\mathrm{s}^i(&o_t^i,a_t^i,\{z_t^j\}_{j\neq i})\nonumber \\
    &=\ r_\mathrm{ext}+r_\mathrm{nov}^i(o_t^i,a_t^i)+\lambda r_\mathrm{hin}^i(o_t^i,a_t^i,\{z_t^j\}_{j\neq i})\nonumber \\
    &=\ r_\mathrm{ext}+\sum_{j}u^j_t+\lambda\sum_{j\neq i}z_t^j\log\frac{p(a_t^i | o_t^i,z_t^j)}{\pi^i(a_t^i | o_t^i)},
\end{align}
where $\lambda$ is a hyperparameter that denotes the weight of the hindsight-based intrinsic reward. In other words, $\lambda$ controls the weight between encouraging agents to visit globally novel states and encouraging agents to influence other agents' exploration. Since the calculation of the hindsight-based intrinsic reward requires on-policy samples, we use independent PPO (IPPO) \cite{schulman2017proximal,de2020independent} as the base RL algorithm and train each agent $i$ with shaped reward \eqref{equ:final_r}. \cref{alg:mace} summarizes our method from the perspective of an individual agent $i$. To guarantee scalability, we also propose a scalable hindsight-based intrinsic reward using weighted mutual information between the agent's action and the summation of all other agents' accumulated novelty, described in \cref{app:scale}.

\begin{algorithm}[t]
\caption{MACE for each agent $i$}
\label{alg:mace}
\begin{algorithmic}
    \FOR{$e=1$ {\bfseries to} $M$}
    \FOR{$t=1$ {\bfseries to} $T$}
    \STATE Take action $a^i_t\sim \pi^i(\cdot | o^i_t)$
    \STATE Observe reward $r_\mathrm{ext}$ and next observation $o^i_{t+1}$
    \STATE Compute local novelty $u^i_t=\mathrm{novelty}(o^i_{t+1})$
    \STATE Send $u^i_t$ to other agents
    \STATE Receive $\{u^j_t\}_{j\neq i}$ from other agents
    \STATE Collect $(o_t^i,a_t^i,r_\mathrm{ext},\{u_t^j\}_{j=1}^N,o_{t+1}^i)$ into buffer
    \ENDFOR
    \STATE Compute $z_t^j=\sum_{t'=t}^T\gamma^{t'-t}u_{t'}^j$
    \STATE Update estimated distribution $p(a_t^i | o_t^i,z_t^j)$
    \STATE Update policy $\pi^i$ using shaped reward \eqref{equ:final_r} with IPPO
    \ENDFOR
\end{algorithmic}
\end{algorithm}

\section{Related Work}

\textbf{Single-agent exploration.} Advanced RL algorithms have been developed to improve exploration. Providing the agent with a manually designed intrinsic reward has been proven effective in environments with sparse rewards like Montezuma’s Revenge \cite{brockman2016gym}. Typically, the intrinsic reward is set to be the novelty of the state, \textit{e.g.}, the inverse of the visit count: $r_\mathrm{int}(s)=\mathrm{novelty}(s)=1/n(s)$, to encourage the agent to take action towards states it seldom visits. However, states in real problems are usually high-dimensional, meaning that $n(s)$ is impossible to count in most cases. Count-based methods solve this problem by introducing pseudo-count \cite{bellemare2016unifying} or hashing to discretize states \cite{tang2017exploration}. Other methods measure novelty from different perspectives, including prediction error of transition model \cite{pathak2017curiosity,burda2018large,kim2019emi}, prediction error of state features \cite{burda2018exploration}, policy discrepancy \cite{flet2020adversarially}, state marginal matching \cite{lee2019efficient}, deviation of policy cover \cite{zhang2021made}, uncertainty \cite{houthooft2016vime,pathak2019self}, and TD error of random reward \cite{ramesh2022exploring}. Recent work places an episodic restriction on intrinsic reward, where the intrinsic reward obtained by an agent visiting a repeated state within an episode will be reduced \cite{badia2019never,raileanu2019ride,zhang2021noveld,henaff2022exploration}. 

\vspace{2mm}
\noindent \textbf{Multi-agent exploration.} Exploration in multi-agent environments requires intrinsic reward that is different from that in single-agent environments. \citet{iqbal2019coordinated} proposed several types of intrinsic reward which take into consideration the novelty of agent $i$'s observation from the perspective of agent $j$. 
EITI/EDTI \cite{wang2019influence} focuses on encouraging the agent to states or observations where the agent influences other agents' transition or value function. EMC \cite{zheng2021episodic} uses the summation of the prediction errors of local Q-functions as the shared intrinsic reward. MAVEN \cite{mahajan2019maven} improves multi-agent exploration by maximizing the mutual information between the trajectory and a latent variable, by which agents are encouraged to visit diverse trajectories. CMAE \cite{liu2021cooperative} combines the goal-based method with a state space dimension selection technique to adapt to the exponentially increased state space. MASER \cite{jeon2022maser} selects goals from the observation space instead of the state space.

However, these methods follow the CTDE setting, where unlimited extra information can be used to ease training. Some use QMIX \cite{rashid2018qmix} as their backbone \cite{mahajan2019maven,zheng2021episodic,liu2021cooperative,jeon2022maser}, while others require agents to share their local observations and actions \cite{iqbal2019coordinated,wang2019influence}.
In contrast, MACE is built on top of decentralized learning algorithms and requires neither a centralized Q-function like QMIX nor the communication of observations and actions between agents. It only needs to pass a floating point number, \textit{i.e.}, the local novelty, between agents, resulting in much less communication overhead than the methods mentioned above. Thus, comparison with these multi-agent exploration methods is out of the focus of this paper.

\vspace{2mm}
\noindent \textbf{Mutual information in MARL.} Mutual information is a widely used mathematical tool in MARL. It serves as a measure of correlation between variables, and maximizing or minimizing it can be used as an auxiliary task to improve the performance of MARL algorithms. These algorithms optimize mutual information between different variables to address various aspects of MARL problems, including coordination \cite{jaques2019social,kim2020maximum,cao2021linda,li2022pmic}, diversity \cite{mahajan2019maven,jiang2021emergence,li2021celebrating}, communication \cite{wang2020learning}, and exploration \cite{wang2019influence,mahajan2019maven}. In practice, optimizing mutual information can be achieved by transforming it into an intrinsic reward that is added to the environmental reward \cite{jaques2019social,wang2019influence,jiang2021emergence,li2021celebrating,li2022pmic} or by using it as a regularizer in the overall optimization objective \cite{mahajan2019maven,wang2020roma,kim2020maximum,wang2020learning,cao2021linda}. MACE optimizes a variant of mutual information, namely weighted mutual information, and ensures that it is tractable in decentralized learning. In \cref{app:mi}, we provide a detailed comparison between these algorithms and MACE.

\vspace{2mm}
\noindent \textbf{Decentralized multi-agent reinforcement learning.} By virtue of the advantages of decentralized learning, \textit{e.g.}, easy to implement, better scalability, and more robust \citep{jiang2022i2q}, decentralized learning has attracted much attention from the MARL community. 
The convergence of decentralized learning was theoretically studied for cooperative games in networked settings \cite{zhang2018fully} and for fully decentralized (without communication) stochastic games in tabular cases \cite{Jin2021VLearningA,Daskalakis2022TheCO}, laying the theoretical foundation for decentralized learning. 
\citet{de2020independent,papoudakis2021benchmarking} showed the promising empirical performance of fully decentralized algorithms including IPPO and independent Q-learning (IQL) \citep{tan1993multi} in several cooperative multi-agent benchmarks. 
Recently, \citet{jiang2022i2q} proposed I2Q, a practical fully decentralized algorithm based on Q-learning for cooperative tasks, and proved the convergence of the optimal joint policy, yet limited to deterministic environments. However, the existing work does not take into consideration coordinated exploration and simply uses $\epsilon$-greedy or sampling from the stochastic policy at individual agents. We take a step further to consider decentralized coordinated exploration and thus enable decentralized learning algorithms to solve sparse-reward tasks. As discussed before, our proposed hindsight-based intrinsic reward is more suitable for on-policy algorithms, thus we currently build MACE on IPPO. Combining MACE with off-policy decentralized algorithms like IQL or I2Q is left as future work.

\section{Experiments}
\label{sec:experiment}

In experiments, we evaluate MACE in three environments: GridWorld, Overcooked \cite{carroll2019utility}, and SMAC \cite{samvelyan2019starcraft}. We set all environments sparse-reward. Since we consider decentralized learning, agents in the experiments \textit{do not share their parameters} and learn independently, following existing work \citep{jiang2022i2q}.

\subsection{Setup}

\begin{figure}[t]
    \centering
    \subfigure[\texttt{Pass}]{
        \includegraphics[width=0.3\columnwidth]{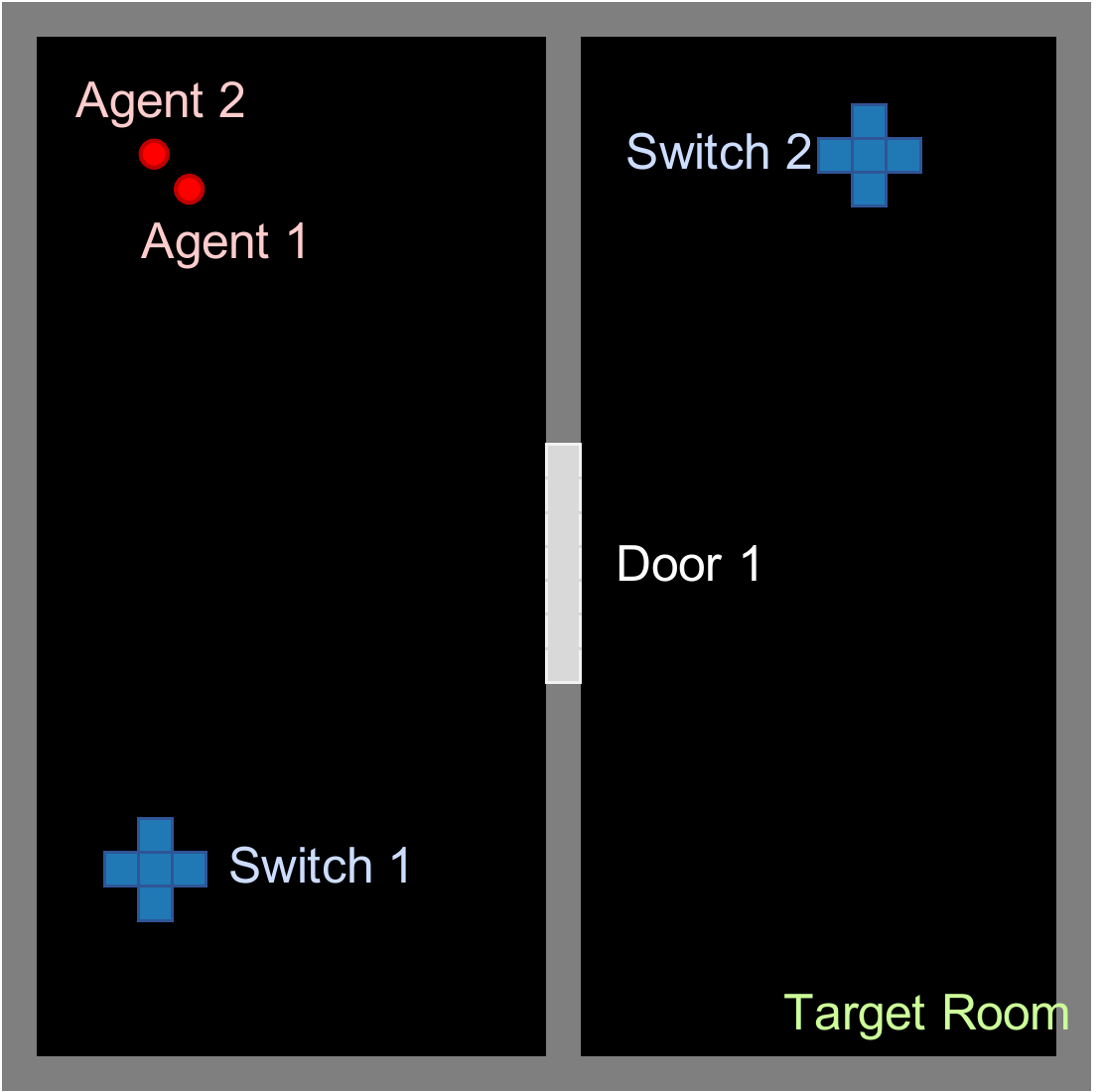}
    }
    \subfigure[\texttt{SecretRoom}]{
        \includegraphics[width=0.3\columnwidth]{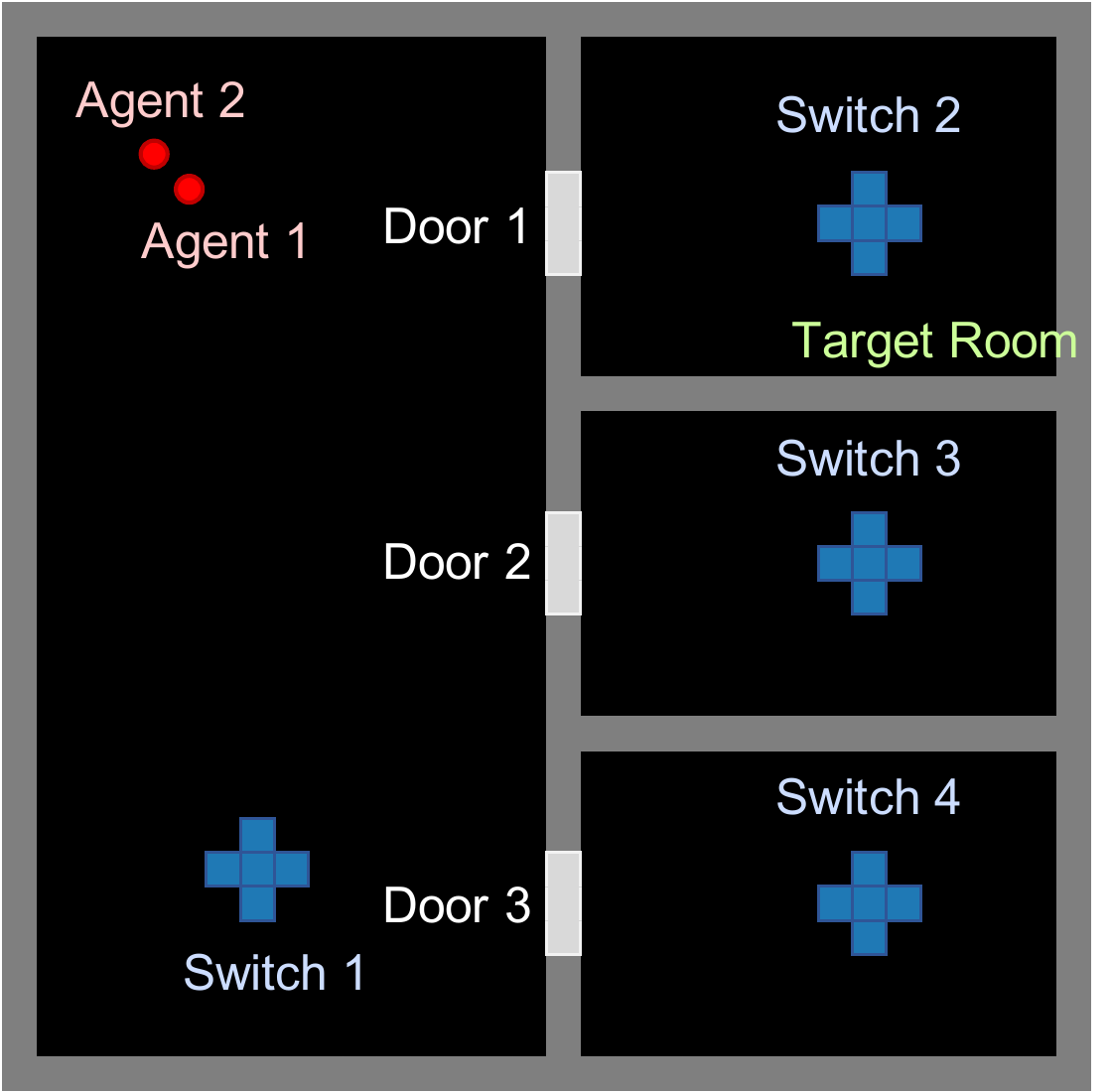}
    }
    \subfigure[\texttt{MultiRoom}]{
        \includegraphics[width=0.3\columnwidth]{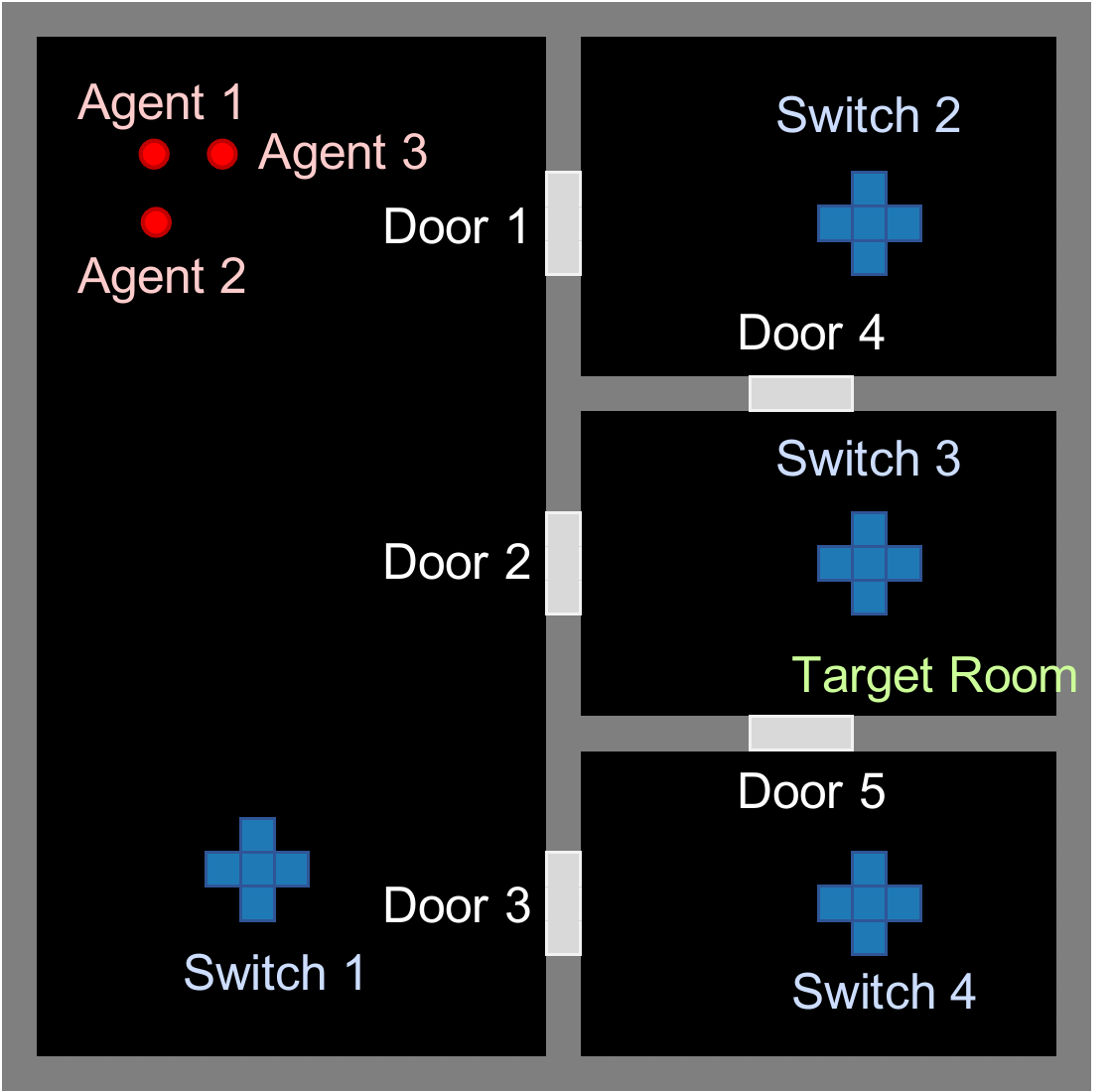}
    }
    \caption{GridWorld: (a) \texttt{Pass}. (b) \texttt{SecretRoom}. (c) \texttt{MultiRoom}.}
    \label{fig:gridworld}
\end{figure}

\vspace{1mm}
\noindent \textbf{GridWorld.} We design three tasks in GridWorld including \texttt{Pass}, \texttt{SecretRoom}, and \texttt{MultiRoom}. \texttt{Pass} and \texttt{SecretRoom} reference tasks in \citet{wang2019influence} and \citet{liu2021cooperative}. In \texttt{MultiRoom}, the task extends to three agents. The goal of all tasks is that all agents enter the target room shown in \cref{fig:gridworld}.

\vspace{1mm}
\noindent \texttt{Pass}: There are two agents in the 30$\times$30 grid. \textit{Door} 1 will open when any switch is occupied. To achieve the goal, one agent needs to reach \textit{switch} 1 to open \textit{door} 1 so that the other agent can enter the target room, then the latter agent needs to reach \textit{switch} 2 to let the former agent come in. 

\vspace{1mm}
\noindent \texttt{SecretRoom}: There are two agents in the 30$\times$30 grid. \textit{Door} $k$ will open when \textit{switch} $k+1$ is occupied and all \textit{doors} will open when \textit{switch} 1 is occupied. Agents need to take the same steps as that in \texttt{Pass} to finish the task. \texttt{SecretRoom} is harder than \texttt{Pass} because there are three rooms on the right to explore but only one room is the target. 

\vspace{1mm}
\noindent \texttt{\texttt{MultiRoom}}: There are three agents in the 30$\times$30 grid. In detail, \textit{door} 1 will open when \textit{switch} 1 is occupied; \textit{door} 3 will open when \textit{switch} 2 is occupied; \textit{door} 2 will open when \textit{switch} 4 is occupied; \textit{door} 4 and \textit{door} 5 will open when \textit{switch} 3 is occupied. More complicated coordinated exploration is required among the three agents. A elaborated description of the solution to this task is provided in \cref{app:gw_detail}.
\vspace{1mm}

The episode ends when all agents are in the target room, and each agent receives a +100 reward. Each agent observes its own location $(x,y)$ and the open states of doors. More details about this environment are available in \cref{app:gw_detail}. These GridWorld tasks serve as didactic examples because the critical states in which exploration of agents interact with each other are obvious, namely the switch locations.

\begin{figure}[t]
    \centering
    \subfigure[\texttt{Base}]{
        \includegraphics[width=0.27\columnwidth]{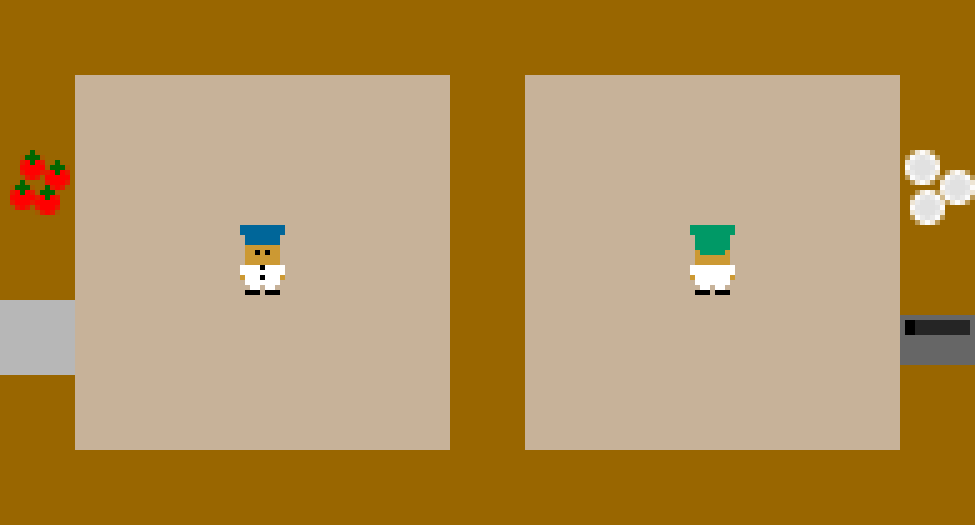}
    }
    \hspace{-5pt}
    \subfigure[\texttt{Narrow}]{
        \includegraphics[width=0.27\columnwidth]{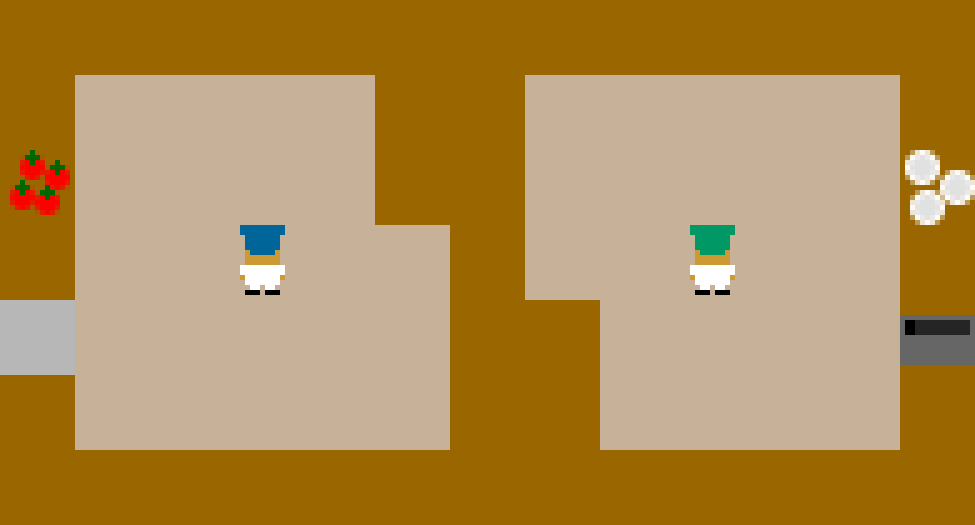}
    }
    \hspace{-5pt}
    \subfigure[\texttt{Large}]{
        \includegraphics[width=0.40\columnwidth]{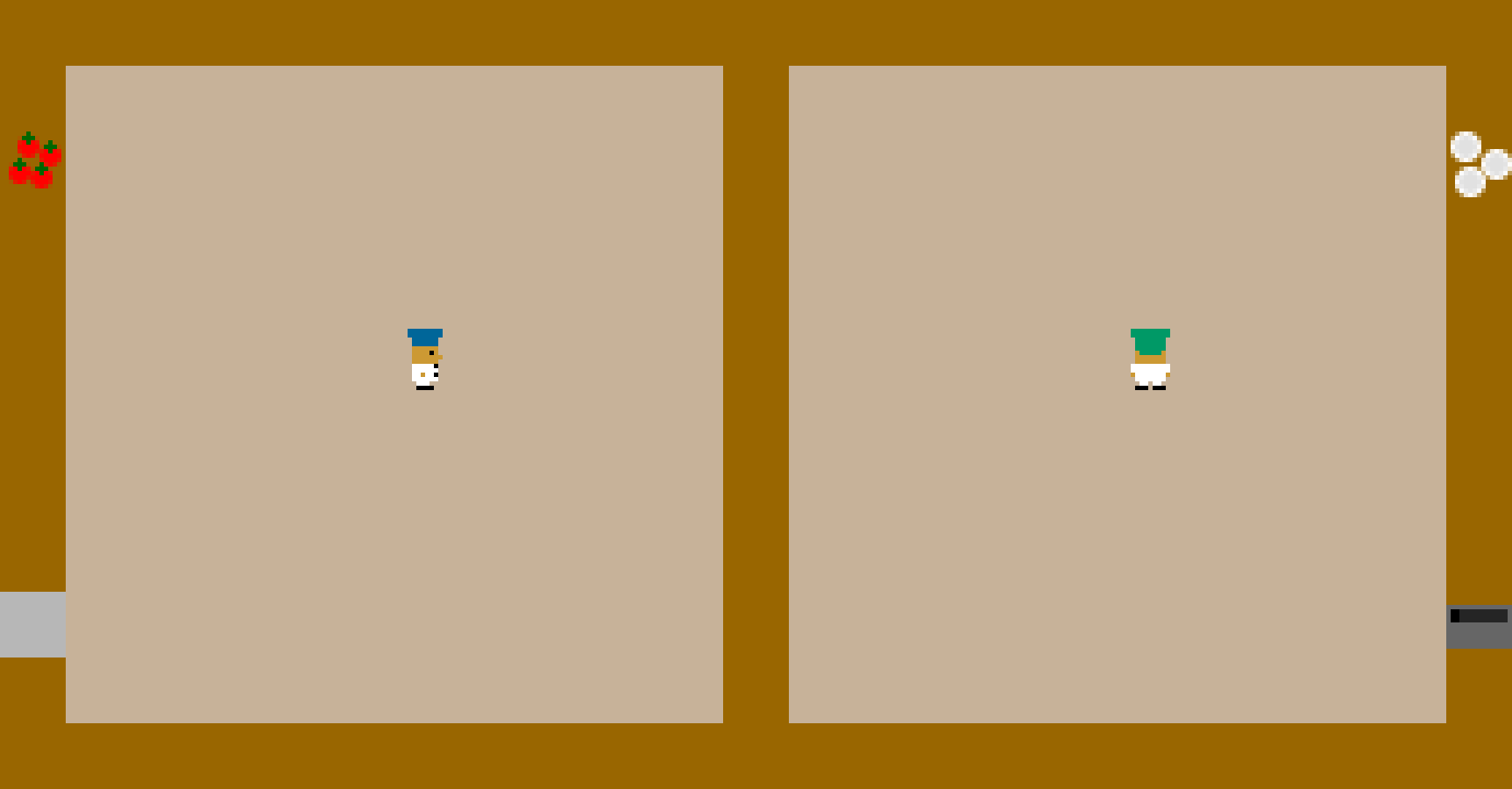}
    }
    \caption{Overcooked: (a) \texttt{Base}. (b) \texttt{Narrow}. (c) \texttt{Large}.}
    \label{fig:overcooked}
\end{figure}

\vspace{2mm}
\noindent \textbf{Overcooked.} We design three tasks in Overcooked \cite{carroll2019utility}: \texttt{Base}, \texttt{Narrow}, and \texttt{Large}. All tasks contain two agents, separated by an impassable kitchen counter as shown in \cref{fig:overcooked}. Therefore, the two agents must cooperate to complete the task. The left agent has access to tomatoes and the serving area (the gray patch in \cref{fig:overcooked}), and the right agent has access to dishes and the pot. Agents need to put one tomato into the pot, cook it, put the resulting soup into a dish, and serve it in order by passing items through the counter. When the soup is served, the episode ends, and each agent receives a +100 reward. Compared to \texttt{Base}, \texttt{Narrow} limits the area where items can be passed to only the middle of the counter, and \texttt{Large} increases the size of the entire environment. More details about this environment are available in \cref{app:oc_detail}.

\vspace{2mm}
\noindent \textbf{SMAC.} We use three maps in SMAC \cite{samvelyan2019starcraft} 2.4.10: \texttt{2m\_vs\_1z}, \texttt{3m}, and \texttt{8m}, customized to be sparse-reward. Agents receive a +200 reward if they win the game. In \texttt{3m} and \texttt{8m}, agents also receive a +10 reward when one enemy dies so as to ease the task. In \texttt{8m}, we use the scalable hindsight-based intrinsic reward described in \cref{app:scale}.

\vspace{2mm}
\noindent \textbf{Implementation.} For all tasks, we implement PPO leveraging GRU \cite{cho2014properties} as the policy and critic function. In GridWorld, given that the observation space is small and discrete, we use the inverse of visit counts as the novelty measurement and use a table to record each observation's visit count. Also, we use a table to record recent discretized accumulated novelty $z_t^j$ and corresponding observation $o_t^i$ and action $a_t^i$. Then we can estimate the posterior distribution $p(a_t^i | o_t^i,z_t^j)$ from the table. In Overcooked and SMAC, we use RND \cite{burda2018exploration} as the novelty measurement and use an MLP to fit the posterior distribution $p(a_t^i | o_t^i,z_t^j)$ via supervised learning. More details about the novelty measurement, the estimation of the posterior distribution, and the hyperparameters are available in \cref{app:imp}.

\subsection{GridWorld}

\begin{figure*}
    \centering
    \subfigure[\texttt{Pass}]{
        \includegraphics[width=0.55\columnwidth]{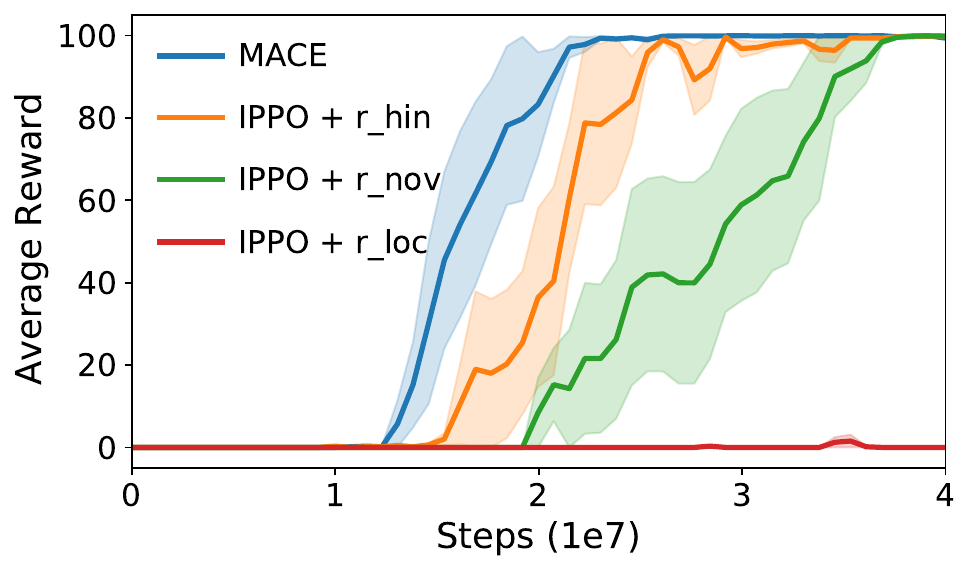}
    }
    \qquad
    \subfigure[\texttt{SecretRoom}]{
        \includegraphics[width=0.55\columnwidth]{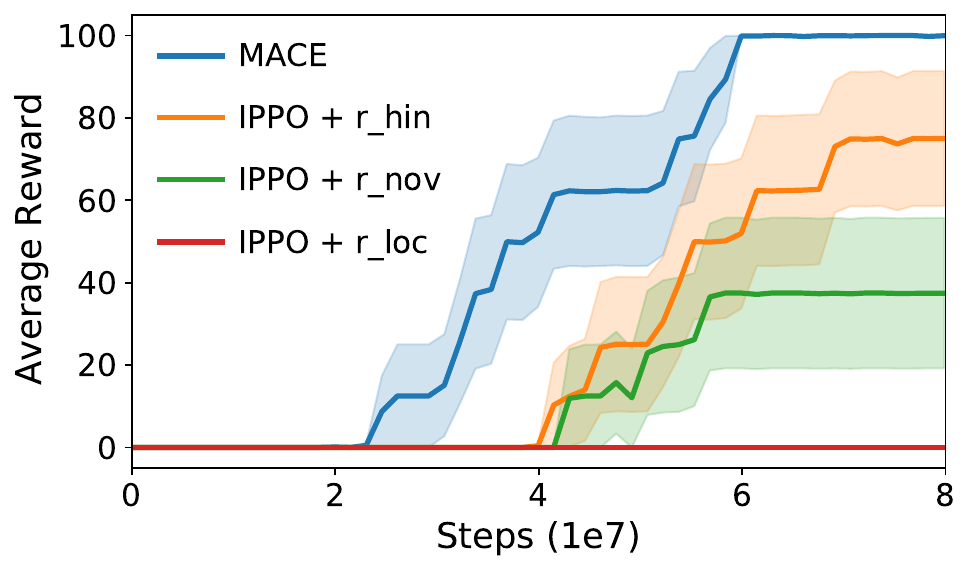}
    }
    \qquad
    \subfigure[\texttt{MultiRoom}]{
        \includegraphics[width=0.55\columnwidth]{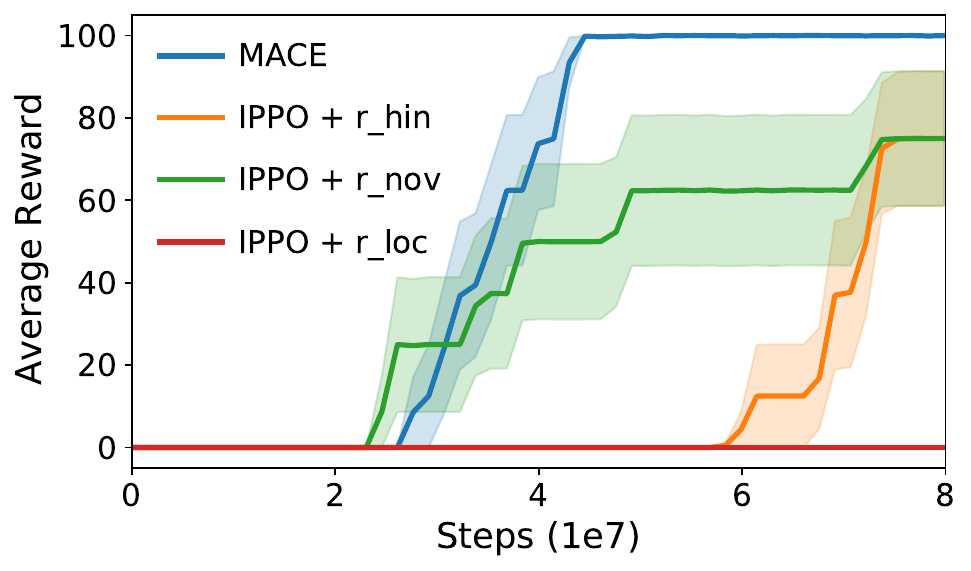}
    }
    \caption{Learning curves of MACE compared with IPPO+r\_loc, IPPO+r\_nov, and IPPO+r\_hin on three GridWorld tasks: (a) \texttt{Pass}, (b) \texttt{SecretRoom}, and (c) \texttt{MultiRoom}.}
    \label{fig:gridworld_base}
\end{figure*}

\begin{figure*}[t]
    \centering
    \subfigure[\texttt{Pass}]{
        \includegraphics[width=0.55\columnwidth]{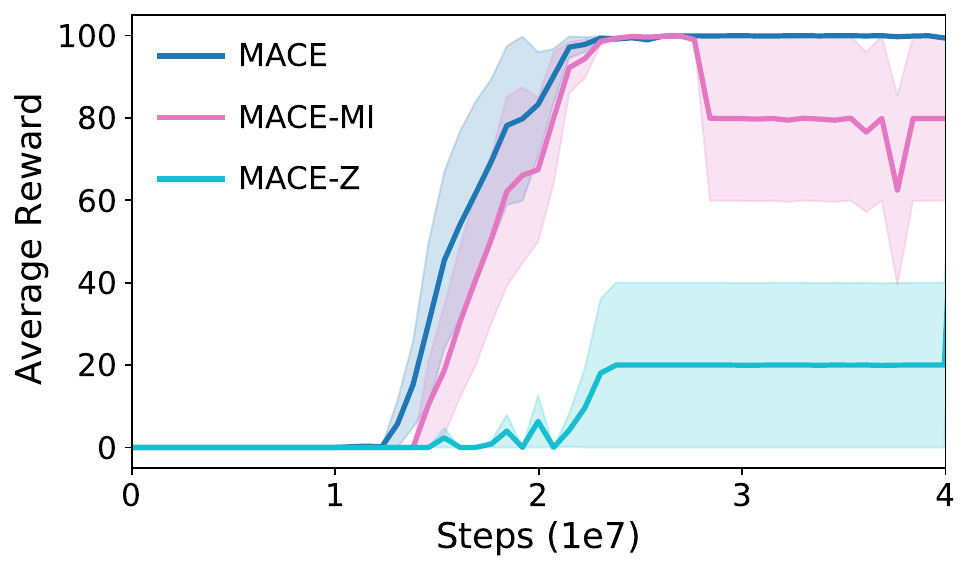}
    }
    \qquad
    \subfigure[\texttt{SecretRoom}]{
        \includegraphics[width=0.55\columnwidth]{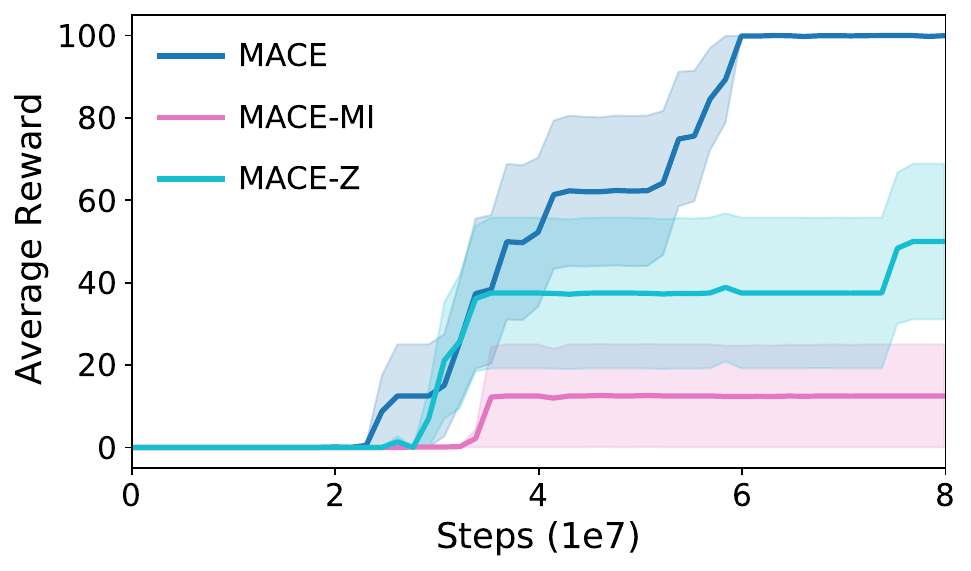}
    }
    \qquad
    \subfigure[\texttt{MultiRoom}]{
        \includegraphics[width=0.55\columnwidth]{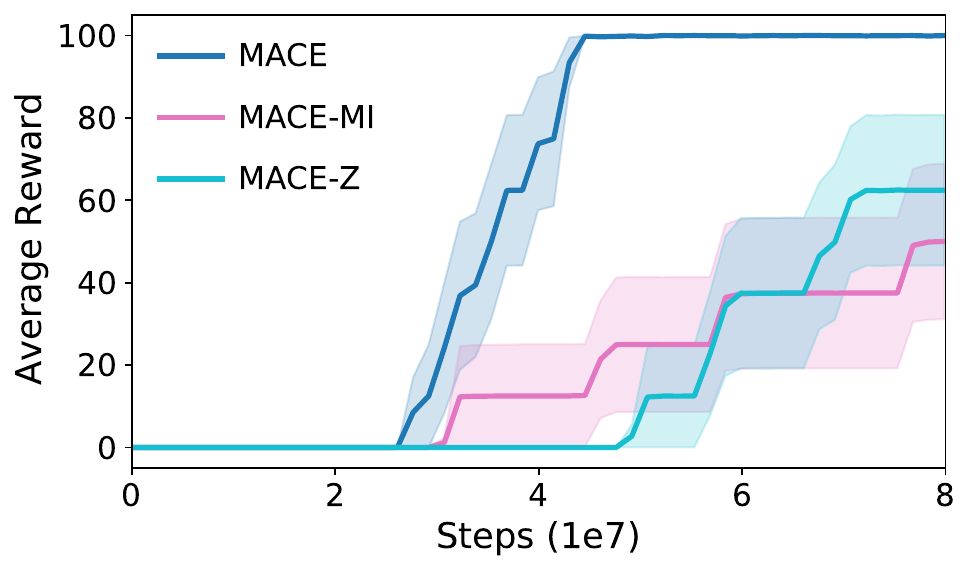}
    }
    \caption{Learning curves of MACE compared with MACE-MI and MACE-Z on three GridWorld tasks: (a) \texttt{Pass}, (b) \texttt{SecretRoom}, and (c) \texttt{MultiRoom}. }
    \label{fig:gridworld_ablation}
\end{figure*}

We first verify the effectiveness of MACE in promoting coordinated exploration by ablation studies. We compare MACE with the following methods:
\begin{itemize}[itemsep=1pt]
    \item \textbf{IPPO+r\_loc}: Agents are trained with $r_\mathrm{ext}+u_t^i$, meaning that they only take into consideration the local novelty.
    \item \textbf{IPPO+r\_nov}: Agents are trained with $r_\mathrm{ext}+r_\mathrm{nov}^i$, meaning that they explore via approximated global novelty.
    \item \textbf{IPPO+r\_hin}: Agents are trained with $r_\mathrm{ext}+u_t^i+\lambda r_\mathrm{hin}^i$, meaning that they explore via local novelty and influence on other agents' exploration. $\lambda$ here keeps the same as that used in MACE.
\end{itemize}

The results are shown in \cref{fig:gridworld_base}. Each curve shows the mean reward of several runs with different random seeds (5 runs in \texttt{Pass}, 8 runs in \texttt{SecretRoom} and \texttt{MultiRoom}) and shaded regions indicate standard error. IPPO+r\_loc is unable to solve any task because the local novelty is unreliable and insufficient for coordinated exploration. IPPO+r\_nov performs better than IPPO+r\_loc, indicating that taking into account the local novelty of other agents to approximate the global novelty is helpful for coordinated exploration. MACE achieves the best performance on all three tasks, suggesting that the hindsight-based intrinsic reward can further boost coordinated exploration by finding the critical states where the agent influences other agents' exploration. This can also be evidenced by the fact that IPPO+r\_hin achieves higher rewards than IPPO+r\_loc. We also conduct a parameter study on $\lambda$, available in \cref{app:lambda}. A comparison with additional baselines, such as IPPO, can be found in \cref{app:ctde}.

\begin{figure}[t]
    \centering
    \subfigure[novelty-based]{
        \includegraphics[width=0.35\columnwidth]{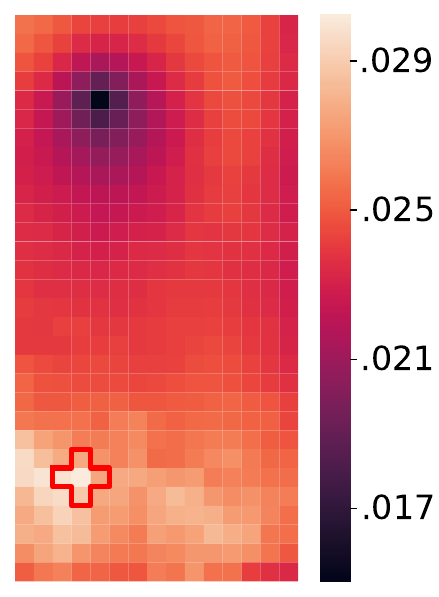}
        \label{fig:ir_vis_nov}
    }
    \quad
    \subfigure[hindsight-based]{
        \includegraphics[width=0.35\columnwidth]{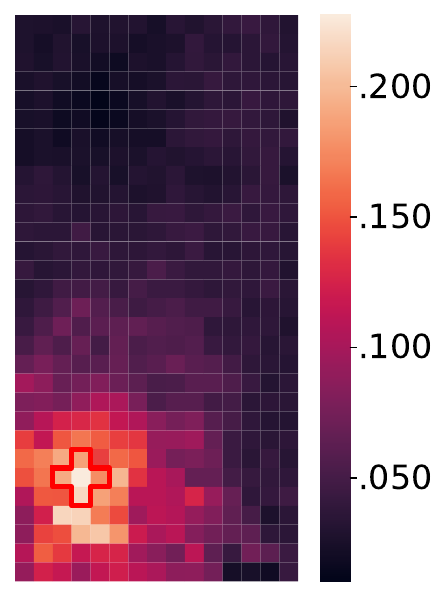}
        \label{fig:ir_vis_inf} 
    }
    \caption{Visualization of the averaged (a) novelty-based intrinsic reward and (b) hindsight-based intrinsic reward received by agent 1 at different positions. The maps only show the intrinsic rewards in the left room in \texttt{Pass}. The red cross highlights the location of \textit{switch} 1.}
    \label{fig:ir_vis}
\end{figure}

\vspace{2mm}
\noindent \textbf{Reward visualization.} To further illustrate how the intrinsic rewards work, we visualize the novelty-based intrinsic reward and the hindsight-based intrinsic reward of agent 1 in the left room in \texttt{Pass}, averaging over 700 to 1000 PPO updates. The critical states consist of one agent stepping on one switch because it will open the middle door and allow the other agent to enter the target room and explore. As shown in \cref{fig:ir_vis_nov}, agent 1 earns higher novelty-based intrinsic rewards at the bottom of the left room than at the top. This is because, after agent 1 steps on \textit{switch} 1 and agent 2 enters the target room successfully, agent 1 will receive high local novelty $u^2$ from agent 2 in the following timesteps. It is noticeable that the hindsight-based intrinsic reward can locate the critical states more accurately. As shown in \cref{fig:ir_vis_inf}, agent 1 earns the highest hindsight-based intrinsic reward around \textit{switch} 1, because these positions correspond to high weighted mutual information: If agent 1 moves towards \textit{switch} 1, it is likely for agent 1 to eventually step on \textit{switch} 1 and allow agent 2 to enter the target room with high novelty; If agent 1 moves away from \textit{switch} 1, agent 2 may be blocked from the target room.

\begin{figure*}[t]
    \centering
    \subfigure[\texttt{Base}]{
        \includegraphics[width=0.55\columnwidth]{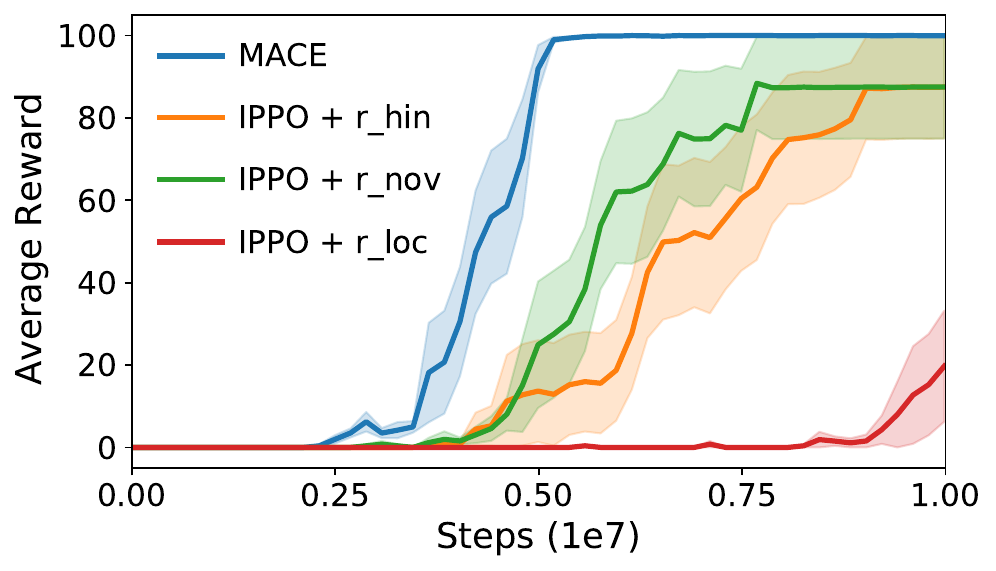}
    }
    \qquad
    \subfigure[\texttt{Narrow}]{
        \includegraphics[width=0.55\columnwidth]{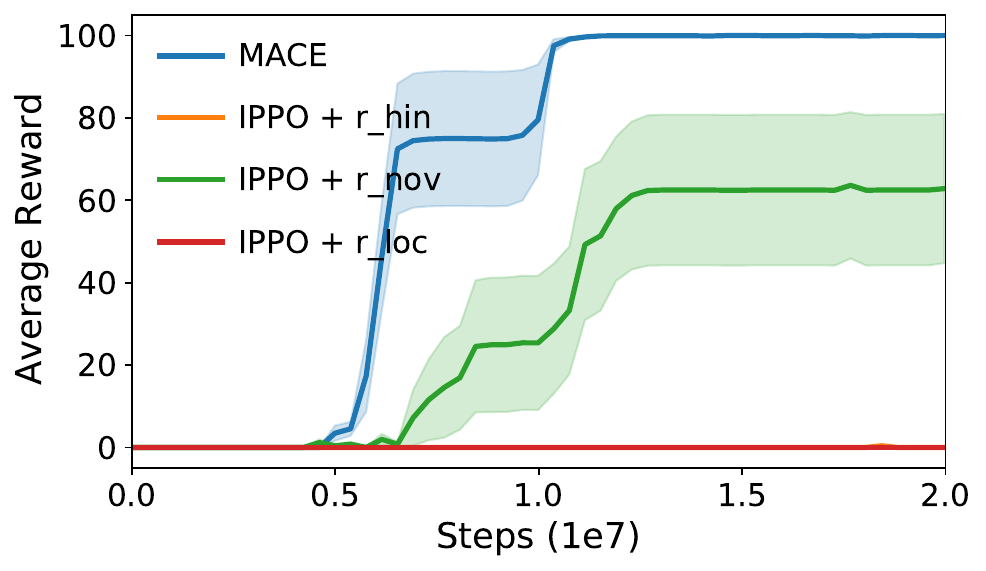}
    }
    \qquad
    \subfigure[\texttt{Large}]{
        \includegraphics[width=0.55\columnwidth]{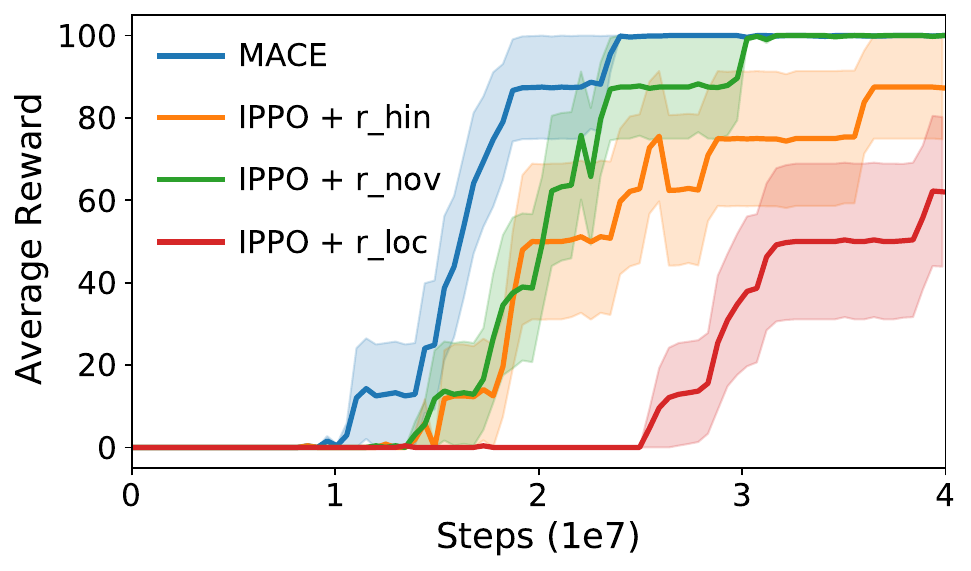}
    }
    \caption{Learning curves of MACE compared with IPPO+r\_loc, IPPO+r\_nov, and IPPO+r\_hin on three Overcooked tasks: (a) \texttt{Base}, (b) \texttt{Narrow}, and (c) \texttt{Large}. }
    \label{fig:overcooked_result}
\end{figure*}

\begin{figure*}[t]
    \centering
    \subfigure[\texttt{2m\_vs\_1z}]{
        \includegraphics[width=0.55\columnwidth]{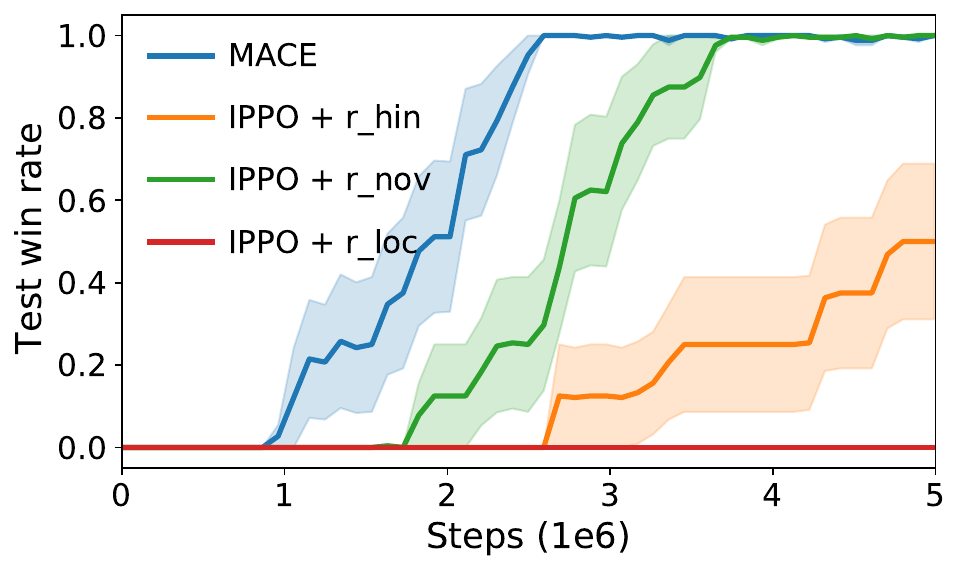}
    }
    \qquad
    \subfigure[\texttt{3m}]{
        \includegraphics[width=0.55\columnwidth]{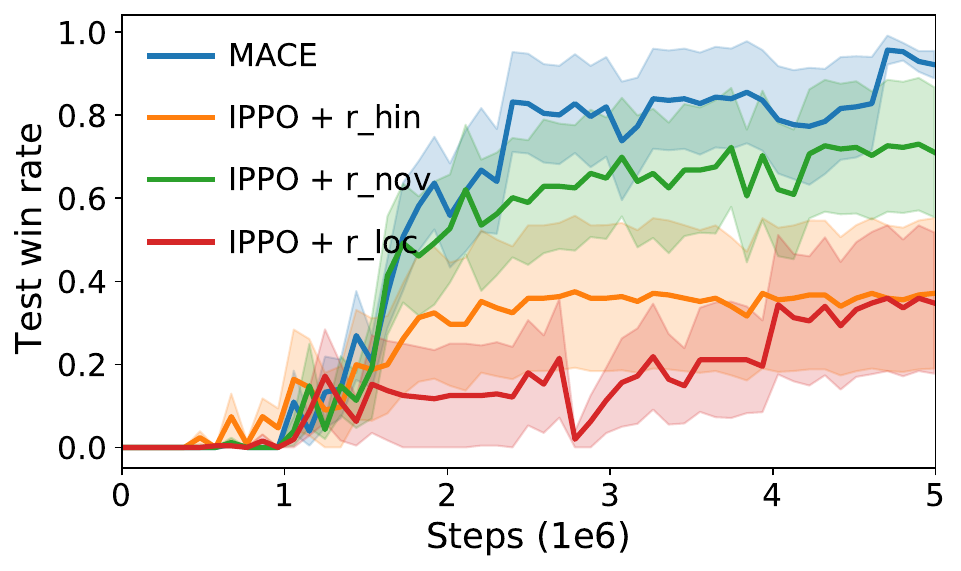}
    }
    \qquad
    \subfigure[\texttt{8m}]{
        \includegraphics[width=0.55\columnwidth]{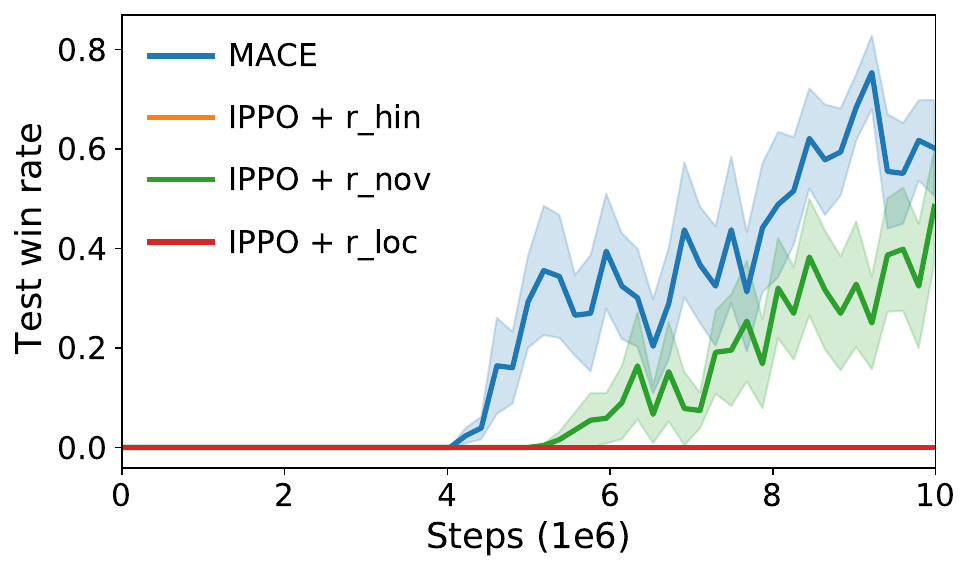}
    }
    \caption{Learning curves of MACE compared with IPPO+r\_loc, IPPO+r\_nov, and IPPO+r\_hin on three SMAC maps: (a) \texttt{2m\_vs\_1z}, (b) \texttt{3m}, and (c) \texttt{8m}. }
    \label{fig:sc_result}
\end{figure*}

\vspace{2mm}
\noindent \textbf{Ablation.} The hindsight-based intrinsic reward \eqref{equ:final_rij} consists of two parts: $z_t^j$, the accumulated novelty of agent $j$, and a logarithmic term $\log\frac{p(a_t^i | o_t^i,z_t^j)}{\pi^i(a_t^i | o_t^i)}$. We test the effectiveness of the two parts separately to verify that none of them alone leads to MACE's high performance. MACE-MI replaces the hindsight-based intrinsic reward with the logarithmic term. We name it `MI' because the expectation of this term equals the mutual information between $a_t^i$ and $z_t^j$ given $o_t^i$. The results in \cref{fig:gridworld_ablation} show that MACE-MI is less effective than MACE in all tasks, validating our claim in \cref{sec:3_2} that weighted mutual information is a more effective measure of the influence on other agent's exploration than mutual information. MACE-Z, replacing the hindsight-based intrinsic reward with $z_t^j$, also performs worse than MACE, indicating that utilizing other agents' accumulated novelty as the intrinsic reward, regardless of whether it is related to the agent's own actions, is ineffective.

\subsection{Overcooked}

We then evaluate the performance of MACE in Overcooked \cite{carroll2019utility} and compare it with IPPO+r\_loc, IPPO+r\_nov, and IPPO+r\_hin. The results are shown in \cref{fig:overcooked_result}. Each curve shows the mean reward of 8 runs with different random seeds, and shaded regions indicate standard error. MACE outperforms others, proving that MACE also works in the high-dimensional state space where the novelty is calculated via RND \cite{burda2018exploration} and the posterior distribution $p(a_t^i | o_t^i,z_t^j)$ is learned via supervised learning. We also observe that IPPO+r\_nov outperforms IPPO+r\_hin in all tasks, especially in \texttt{Narrow}, suggesting that the novelty-based intrinsic reward may play a more critical role in such complicated tasks.

\subsection{SMAC}

We further examine MACE in more complex SMAC \cite{carroll2019utility} tasks and compare it with IPPO+r\_loc, IPPO+r\_nov, and IPPO+r\_hin. The results are shown in \cref{fig:sc_result} for the three maps: \texttt{2m\_vs\_1z}, \texttt{3m}, and \texttt{8m}. Each curve shows the mean reward of 8 runs with different random seeds, and shaded regions indicate standard error. MACE learns faster or achieves a higher win rate than the other three methods, by which we verify the effectiveness of MACE on sparse-reward tasks in such a high-dimensional complex environment. In addition, we demonstrate that MACE could also work on some SMAC tasks with the dense reward, as shown in \cref{app:dense}.

\section{Conclusion}

We propose MACE to enable multi-agent coordinated exploration in decentralized learning with limited communication. MACE uses a novelty-based intrinsic reward and a hindsight-based intrinsic reward to guide exploration. The former is devised to narrow the gap between the local novelty and the unavailable global novelty. The latter is designed to find the critical states where one agent's action influences other agents' exploration, measured by the newly introduced weighted mutual information metric. Through empirical evaluation, we demonstrate the effectiveness of MACE in a variety of sparse-reward multi-agent tasks which need agents to explore cooperatively. In addition, we acknowledge that certain limitations exist, such as the need for a fully-connected communication network to share novelty among agents. Future work could explore ways to further reduce the number of necessary connections besides the bandwidth of communication channels.

\section*{Acknowledgments}
This work was supported by NSF China under grant 62250068. The authors would like to thank the anonymous reviewers for their valuable comments. 

\bibliography{aaai24}

\newpage
\appendix
\onecolumn

\section{Mutual Information in MARL}
\label{app:mi}

Mutual information has been widely used in multi-agent reinforcement learning. These MARL algorithms differ in terms of the variables used to compute mutual information, the purpose for which mutual information is used, and the form in which mutual information is employed. We provide a summary of MARL algorithms that incorporate mutual information in \cref{tab:mi_sum}. 

\begin{table}[htb]
    \centering
    \caption{Mutual Information in MARL.}
    \renewcommand{\arraystretch}{1.4}
    \vskip 0.1in
    \begin{threeparttable}
    \begin{tabular}{cc >{\centering\arraybackslash}p{0.16\columnwidth}p{0.3\columnwidth}}
        \toprule
        Method & Mutual Information & Purpose & Meaning \\
        \midrule
        SI \cite{jaques2019social} \tnote{*} & $I(a^j_t;a^i_t | s_t)$ & coordination & correlation between agent $i$'s action and agent $j$' action \\
        EITI \cite{wang2019influence} \tnote{*} & $I(s^j_{t+1};s^i_t,a^i_t | s^j_t,a^j_t)$ & exploration & correlation between agent $i$'s action and agent $j$' next state \\
        MAVEN \cite{mahajan2019maven} \tnote{\dag} & $I(\boldsymbol{\tau};z)$ & exploration / diversity & correlation between joint trajectory and latent variable \\
        ROMA \cite{wang2020roma} \tnote{\dag} & $I(\rho^i_t;\tau^i_{t-1}|o^i_t)$ & diversity & correlation between agent $i$'s trajectory and its role \\
        MMI \cite{kim2020maximum} \tnote{\dag} & $I(\pi^i(\cdot|s_t);\pi^j(\cdot|s_t))$ & coordination & correlation between agent $i$'s policy and agent $j$'s policy \\
        NDVF \cite{wang2020learning} \tnote{\dag} & $I(a_t^j;m_t^{i j}|\tau_t^j,m_t^{(-i) j})$ & communication & correlation between agent $j$'s action and message intended from agent $i$ to agent $j$ \\
        EOI \cite{jiang2021emergence} \tnote{*} & $I(o_t^i;i)$ & diversity & correlation between agent $i$'s observation and its index \\
        CDS \cite{li2021celebrating} \tnote{*} & $I(\tau^i_T;i)$ & diversity & correlation between agent $i$'s trajectory and its index \\
        LINDA \cite{cao2021linda} \tnote{\dag} & $I(\boldsymbol{c}^i_j;\tau^j|\tau^i)$ & coordination & correlation between agent $j$'s representation from agent $i$ and agent $j$'s trajectory \\
        PMIC \cite{li2022pmic} \tnote{*} & $I(s_t;\boldsymbol{a}_t)$ & coordination & correlation between state and joint action \\
        MACE (ours) \tnote{*} & $\omega I(a^i_t;z^j_t|o^i_t)$ & exploration & correlation between agent $i$'s action and agent $j$'s accumulated novelty, taking into account the magnitude of agent $j$'s accumulated novelty \\
        \bottomrule
    \end{tabular}
    \begin{tablenotes}
        \item[*] Mutual information is employed as an intrinsic reward.
        \item[\dag] Mutual information is employed as a regularizer.
    \end{tablenotes}
    \end{threeparttable}
    \label{tab:mi_sum}
\end{table}

\section{Environment Details}

\subsection{GridWorld}
\label{app:gw_detail}

\begin{figure}[h]
    \centering
    \includegraphics[width=0.3\columnwidth]{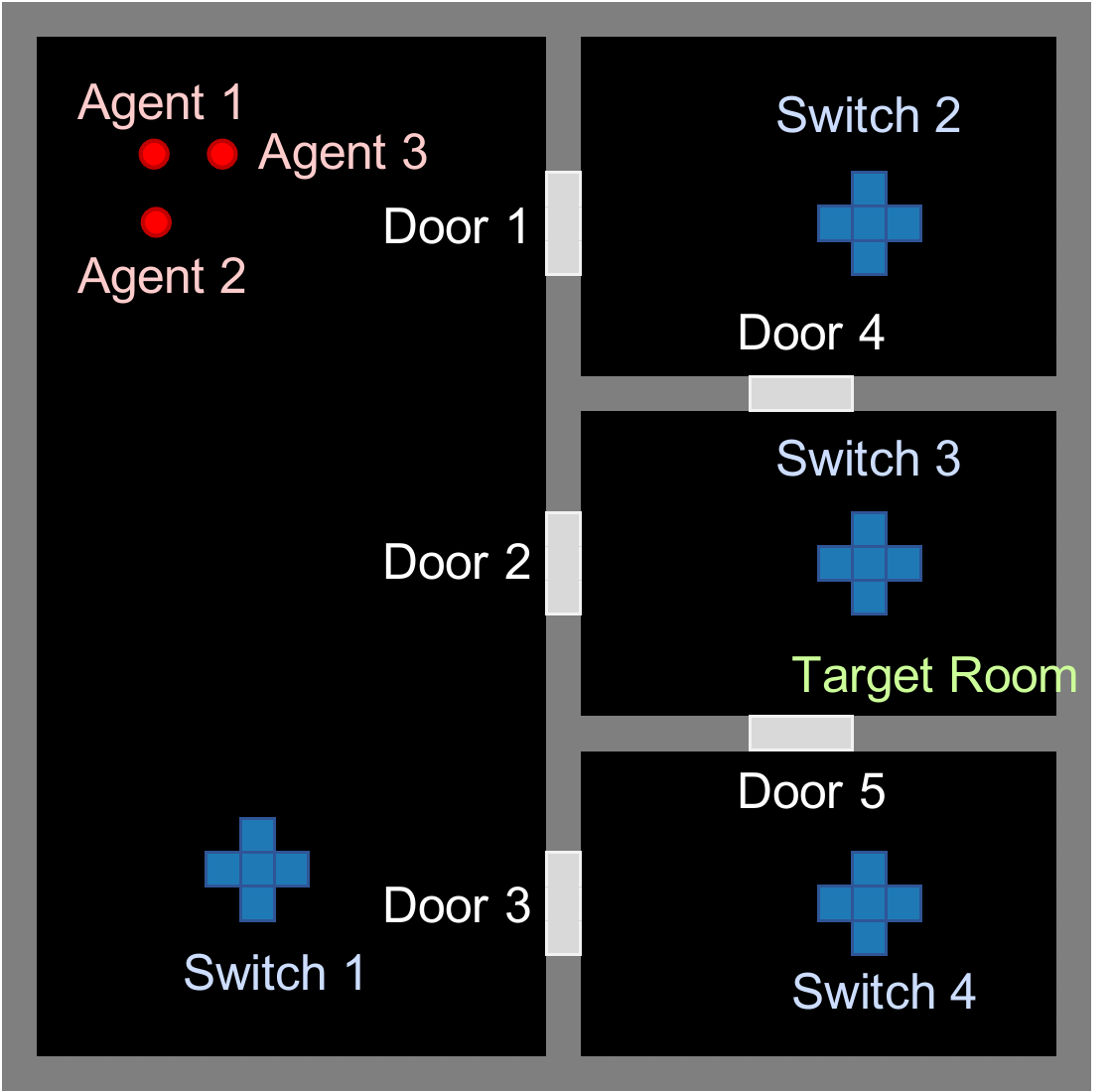}
    \caption{GridWorld environment: \texttt{MultiRoom}}
\end{figure}

The \textit{door-switch} rules in \texttt{MultiRoom} are as follows: 
\begin{itemize}[itemsep=2pt,topsep=2pt,leftmargin=15pt]
    \item \textit{Door} 1 will open when \textit{switch} 1 is occupied.
    \item \textit{Door} 3 will open when \textit{switch} 2 is occupied.
    \item \textit{Door} 2 will open when \textit{switch} 4 is occupied.
    \item \textit{Door} 4 and \textit{door} 5 will open when \textit{switch} 3 is occupied.
\end{itemize}
To achieve the goal, agents need to take the following steps in order: 
\begin{itemize}[itemsep=2pt,topsep=2pt,leftmargin=15pt]
    \item One agent reaches \textit{switch} 1 and lets another agent enter the room containing \textit{switch} 2 through \textit{door} 1.
    \item One agent reaches \textit{switch} 2 and lets another agent enter the room containing \textit{switch} 4 through \textit{door} 3.
    \item One agent reaches \textit{switch} 4 and lets another agent enter the target room through \textit{door} 2.
    \item One agent reaches \textit{switch} 3 and lets the other two agents enter the target room through \textit{door} 4 or \textit{door} 5.
\end{itemize}

In GridWorld, each agent can observe its own location $(x,y)$ and the open states of doors indicated by 0 (closed) and 1 (open). So the dimensions of observation spaces in the three tasks are 3, 5, and 7 respectively. The action space contains four actions, including \textit{move up}, \textit{move down}, \textit{move left}, and \textit{move right}. The maximum episode length is set to 300.
\subsection{Overcooked}
\label{app:oc_detail}

\begin{figure}[htb]
    \centering
    \subfigure[\texttt{Base}]{
        \includegraphics[width=0.3\columnwidth]{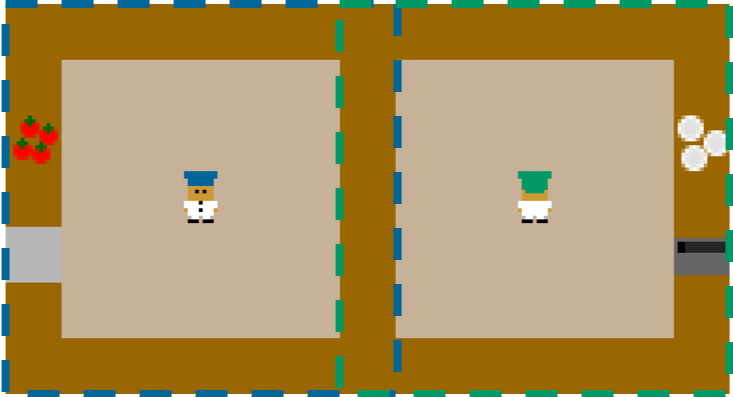}
    }
    \qquad
    \qquad
    \subfigure[\texttt{Narrow}]{
        \includegraphics[width=0.3\columnwidth]{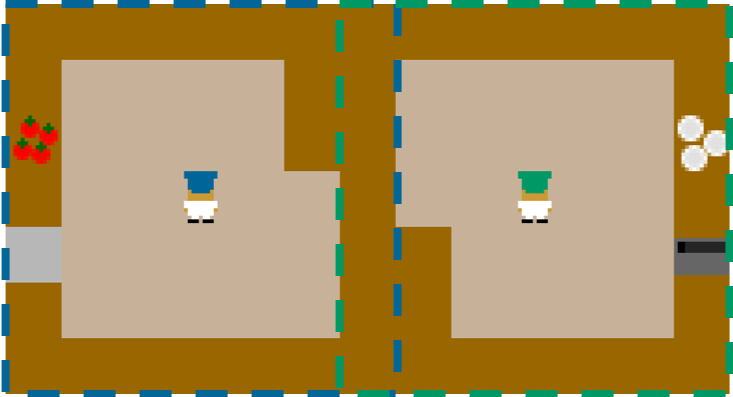}
    }
    \vskip -0.1in
    \caption{Observation ranges in (a) \texttt{Base} and (b) \texttt{Narrow}. The blue dashed box represents the observation range of the left agent, and the green dashed box represents the observation range of the right agent.}
    \label{fig:obs_range}
\end{figure}

All tasks contain two agents, separated by an impassable kitchen counter as shown in \cref{fig:obs_range}. Therefore, the two agents must cooperate to complete the task. The left agent has access to tomatoes and the serving area (the gray patch in \cref{fig:obs_range}), and the right agent has access to dishes and the pot. Agents need to put one tomato into the pot, cook it, put the resulting soup into a dish, and serve it in order by passing items through the counter. We use the open-source Overcooked environment\footnote{\url{https://github.com/HumanCompatibleAI/overcooked_ai}} of \citet{carroll2019utility} (MIT License). We make some modifications to the original environment, including:
\begin{enumerate}[itemsep=2pt,topsep=2pt,leftmargin=15pt]
    \item We restrict the observation ranges of agents. In the original environment, each agent could observe objects regardless of the distance. To increase the bias between the global novelty and the local novelty, we add an option to set the observation range of the agent. Items outside the observation range are treated as non-existent. For example, in our tasks, we could set the left agent to not be able to observe the dishes and the pot in the right room.
    \item We remove the cooking time of the soup to ease the task. In the original environment, it takes 20 timesteps to cook a soup. We set the cost time to 0, \textit{i.e.}, the soup is cooked immediately after the agent interacts with the pot.
    \item We set the episode to end with one successful serving instead of after a fixed timestep. In the original environment, agents need to serve the correct soup in the recipe as many as possible in a fixed-length episode. In our tasks, the only soup in the recipe consists of one tomato, and the episode ends immediately when the correct soup is served.
\end{enumerate}

We use the \textit{featurized\_state} provided by the environment as the observation and add the observation range restriction. The observation space in our tasks contains 38 dimensions. In \texttt{Base} and \texttt{Narrow}, we restrict each agent to observe only the items in its room or on the middle counter, as shown in \cref{fig:obs_range}. We do not restrict agents' observation ranges in \texttt{Large}, because the preliminary experiment shows \texttt{Large} is difficult to learn with the restricted observation range. The action space contains six actions, including \textit{move up}, \textit{move down}, \textit{move left}, \textit{move right}, \textit{interact}, and \textit{stay}. The maximum episode length is set to 300.

\subsection{StarCraft Multi-Agent Challenge}

StarCraft Multi-Agent Challenge (SMAC) \cite{samvelyan2019starcraft} is a commonly used cooperative multi-agent reinforcement learning environment. We use the open-source SMAC environment\footnote{\url{https://github.com/oxwhirl/smac}} (MIT License). We select three maps: \texttt{2m\_vs\_1z}, \texttt{3m}, and \texttt{8m}. In \texttt{2m\_vs\_1z}, two Marines need to defeat an enemy Zealot. In \texttt{3m}, three Marines need to defeat three enemy Marines. In \texttt{8m}, eight Marines need to defeat eight Marines.

\section{Implementation Details}
\label{app:imp}

\subsection{Novelty Estimation}

In GridWorld, the novelty of an observation $o$ is defined as:
\begin{equation*}
    \mathrm{novelty}(o) = 10\cdot\frac{1}{\sqrt{n(x_o,y_o)}},
\end{equation*}
where $n(x_o,y_o)$ is the visit count of the coordinate $(x,y)$ in $o$. Each agent has a $30\times 30$ table to record its visit counts of all coordinates. Each cell in the table records the visit count of the corresponding coordinate.

In Overcooked \cite{carroll2019utility} and SMAC \cite{samvelyan2019starcraft}, we use RND \cite{burda2018exploration} to compute novelty. The novelty is defined as:
\begin{equation}
    \mathrm{novelty}(o) = \left\| f(o;\theta)-\Bar{f}(o) \right\|_2.
    \label{equ:rnd}
\end{equation}
$\Bar{f}(\cdot)$ is a fixed and randomly generated MLP (target network), and $f(\cdot;\theta)$ is an MLP that is trained to minimize \eqref{equ:rnd} (predictor network). Due to training, the more times an observation $o$ is visited, the smaller the difference between the output vectors of these two MLPs on this observation. The network architectures and optimizer settings are available in \cref{tab:rnd_hp}. The predictor network is trained together with IPPO, so other hyperparameters including buffer size, number of mini-batch, and epoch are the same as those in IPPO (\cref{app:ippo_hp}). We do not use reward and observation normalization.

\begin{table}[htb]
    \centering
    \caption{Hyperparameters for RND network.}
    \begin{tabular}{cc}
        \toprule
        Hyperparameter & Value \\
        \midrule
        hidden layers of the target network & [64, 64] \\
        hidden layers of the predictor network & [64, 64, 64, 64] \\
        output size & 32 \\
        optimizer & Adam \\
        optimizer epsilon & 1e-5 \\
        learning rate & 3e-4 \\
        \bottomrule
    \end{tabular}
    \label{tab:rnd_hp}
\end{table}

\subsection{Posterior Distribution Estimation}
\label{app:est_pos}

In GridWorld, we use a count-based method to estimate the posterior distribution $p(a_t^i|o_t^i,z_t^j)$:
\begin{equation}
    \hat{p}(a_t^i|o_t^i,z_t^j) = \frac{n(a_t^i,o_t^i,z_t^j)}{n(o_t^i,z_t^j)},
    \label{equ:est_pos}
\end{equation}
where $n(\cdot)$ is the visit count of each $(a^i,o^i,z^j)$ pair and $(o^i,z^j)$ pair. Each agent $i$ uses a table to record the visit counts of $(a^i,o^i,z^j)$ pairs which are related to agent $j$. $n(o^i,z^j)$ is computed via $\sum_{a^i}n(a^i,o^i,z^j)$. Suppose there are $N$ agents in the environment, each agent requires $N-1$ tables.

To use a table to save the visit counts of all $(a^i,o^i,z^j)$ pairs, we need to discretize $z^j$. However, discretization is faced with a problem: novelty $u^j$ will keep declining with sampling, resulting in a continuous decline of $z^j$, so $z^j$ cannot be divided into bins by setting fixed boundaries. To solve this problem, we first discretize the novelty using percentile. In more detail, after each sampling is completed, we calculate the 20th, 40th, 60th, and 80th percentiles of all $u^j$ in this sampling as bin edges. Then each $u^j$ is divided into one bin and relabelled as 0.1, 0.3, 0.5, 0.7, or 0.9. We denote the relabelled novelty as $\Tilde{u}^j$ and the accumulated relabelled novelty as $\Tilde{z}^j$, where $\Tilde{z}^j_t=\sum_{t'=t}\gamma^{t'-t}\Tilde{u}^j_{t'}$. The scale of $\Tilde{z}^j$ is stable, \textit{i.e.}, between $0.1/(1-\gamma)$ to $0.9/(1-\gamma)$. We replace $z^j$ with $\Tilde{z}^j$ and discretize it into $K$ bins.

In theory, to fulfill \eqref{equ:mc}, we need on-policy samples to calculate \eqref{equ:est_pos}, meaning that $n(\cdot)$ only counts $(a^i,o^i,z^j)$ pairs in current PPO sampling. However, we find that including some off-policy data, \textit{i.e.}, the samples collected in previous PPO sampling, to calculate \eqref{equ:est_pos} can improve the overall performance of the algorithm by variance-bias tradeoff (\cref{app:different_w}). Therefore, we record the samples from the previous $w$ times of PPO sampling and use them to calculate \eqref{equ:est_pos}.

Given the methods described above, the size of a table related to agent $j$ is $w\times |A|\times |S|\times K$. \texttt{Pass}, \texttt{SecretRoom} and \texttt{MultiRoom} share a common $|A|=4$. $|S|$ of three tasks is $30\times 30\times 2$, $30\times 30\times 2^3$, and $30\times 30\times 2^5$, respectively. In our experiments, we set $w=10$ and $K=30$.

In Overcooked \cite{carroll2019utility} and SMAC \cite{samvelyan2019starcraft}, we use an MLP $f_p(\cdot|o,z)$ to estimate each posterior distribution $p(a_t^i|o_t^i,z_t^j)$ via supervised learning. $f_p$ takes as input $o^i$ and $z^j$ and outputs a predicted distribution of action. Then $f_p$ is trained to minimize the cross entropy between the predicted distribution and true action. Given that there are $N-1$ other agents in the environment, each agent requires $N-1$ $f_p$s, each corresponding to one other agent $j$. The common hyperparameters used in Overcooked and SMAC are available in \cref{tab:est_hp}. Like GridWorld, we use another buffer containing previous samples to train these MLPs. In Overcooked, the buffer size is 1e5 for \texttt{Base} and \texttt{Narrow}, and 2e5 for \texttt{Large}. In SMAC, the buffer size is 5e4 for \texttt{2m\_vs\_1z} and 1e4 for \texttt{3m} and \texttt{8m}.

\begin{table}[htb]
    \centering
    \caption{Common hyperparameters for $f_p$ in Overcooked and SMAC.}
    \begin{tabular}{cc}
        \toprule
        Hyperparameter & Value \\
        \midrule
        hidden layers & [64, 64] \\
        optimizer & Adam \\
        optimizer epsilon & 1e-5 \\
        learning rate & 3e-4 \\
        epoch & 40 \\
        num mini-batch & 1 \\
        \bottomrule
    \end{tabular}
    \label{tab:est_hp}
\end{table}

\subsection{IPPO Hyperparameters}
\label{app:ippo_hp}

\cref{tab:common_hp} describes the common hyperparameters for IPPO across all tasks. The meaning of the hyperparameters follows \citet{yu2021surprising}. \cref{tab:specific_hp} shows the specific hyperparameters used in each task, including parameters that control the sampling procedure and $\lambda$ that controls the weight of the hindsight-based intrinsic reward.

\begin{table}[htb]
    \centering
    \caption{Common hyperparameters for IPPO across all tasks.}
    \begin{tabular}{cc}
        \toprule
        Common Hyperparameter & Value \\
        \midrule
        GRU hidden layers & [64] \\
        fc hidden layers & [64, 64] \\
        recurrent data chunk length & 10 \\
        gradient clip norm & 10.0 \\
        gae lambda & 0.95 \\
        gamma & 0.99 \\
        value loss & huber loss \\
        huber delta & 10.0 \\
        num mini-batch & 1 \\
        optimizer & Adam \\
        optimizer epsilon & 1e-5 \\
        actor learning rate & 7e-4 \\
        critic learning rate & 7e-4 \\
        epoch & 10 \\
        activation & ReLU \\
        entropy coef & 0.05 \\
        PPO clip & 0.2 \\
        network initialization & orthogonal \\
        gain & 0.01 \\
        \bottomrule
    \end{tabular}
    \label{tab:common_hp}
\end{table}

\begin{table}[htb]
    \centering
    \caption{Specific hyperparameters for IPPO in each task.}
    \begin{threeparttable}
    \begin{tabular}{cccccccccc}
    \toprule
    \multirow{2}{*}{Hyperparameter} & \multicolumn{3}{c}{GridWorld} & \multicolumn{3}{c}{Overcooked} & \multicolumn{2}{c}{SMAC} \\
     & \texttt{Pass} & \texttt{SR\tnote{*}} & \texttt{MR\tnote{*}} & \texttt{Base} & \texttt{Narrow} & \texttt{Large} & \texttt{2m\_vs\_1z} & \texttt{3m} & \texttt{8m} \\
    \midrule
    num envs & 128 & 128 & 128 & 32 & 32 & 32 & 8 & 8 & 8\\
    buffer length & 300 & 300 & 300 & 300 & 300 & 300 & 600 & 600 & 600 \\
    $\lambda$ in \eqref{equ:final_r} & 0.01 & 0.01 & 0.01 & 0.1 & 0.1 & 0.1 & 1.0 & 0.005 & 0.1 \\
    \bottomrule
    \end{tabular}
    \begin{tablenotes}
        \item[*] \texttt{SR} and \texttt{MR} are short for \texttt{SecretRoom} and \texttt{MultiRoom} respectively.
    \end{tablenotes}
    \end{threeparttable}
    \label{tab:specific_hp}
\end{table}

\section{More Experimental Results}

\subsection{Summation v.s. Maximum of Local Novelty}
\label{app:sum_vs_max}

\begin{figure}[htb]
    \centering
    \includegraphics[width=0.35\columnwidth]{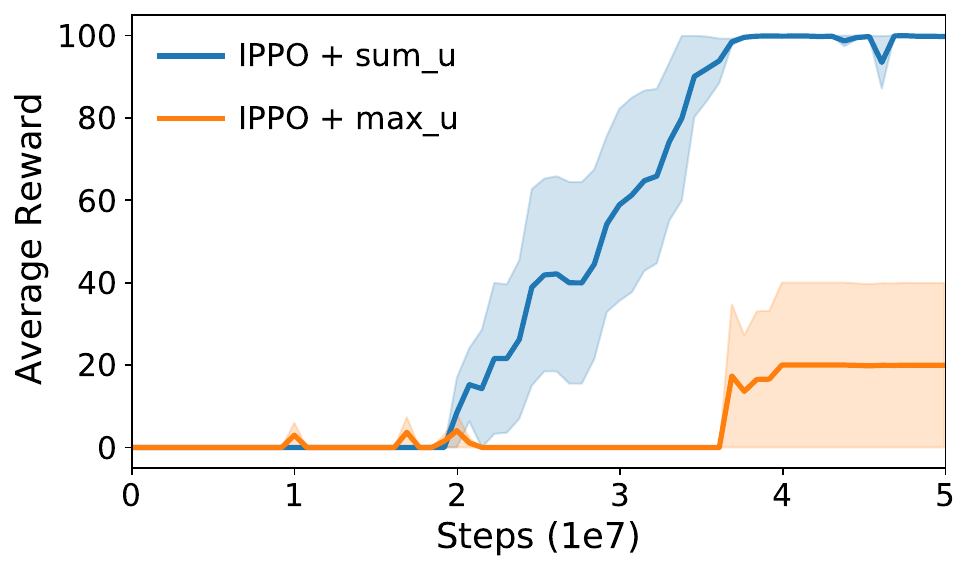}
    \vskip -0.1in
    \caption{Learning curves using summation/maximum of local novelty on \texttt{Pass}.}
    \label{fig:pass_max}
\end{figure}

We compare the performance of IPPO trained with $r_\mathrm{ext}+\sum_ju_t^j$ (denoted as IPPO+sum\_u) and IPPO trained with $r_\mathrm{ext}+\max_ju_t^j$ (denoted as IPPO+max\_u) on \texttt{Pass}. The result in \cref{fig:pass_max} shows that IPPO+sum\_u works better than IPPO+max\_u. Therefore, we choose to use the summation of agents' local novelty as the novelty-based intrinsic reward in MACE instead of the maximum of agents' local novelty.

\subsection{MACE with Different $w$}
\label{app:different_w}

\begin{figure}[htb]
    \centering
    \subfigure[\texttt{Pass}]{
        \includegraphics[width=0.27\columnwidth]{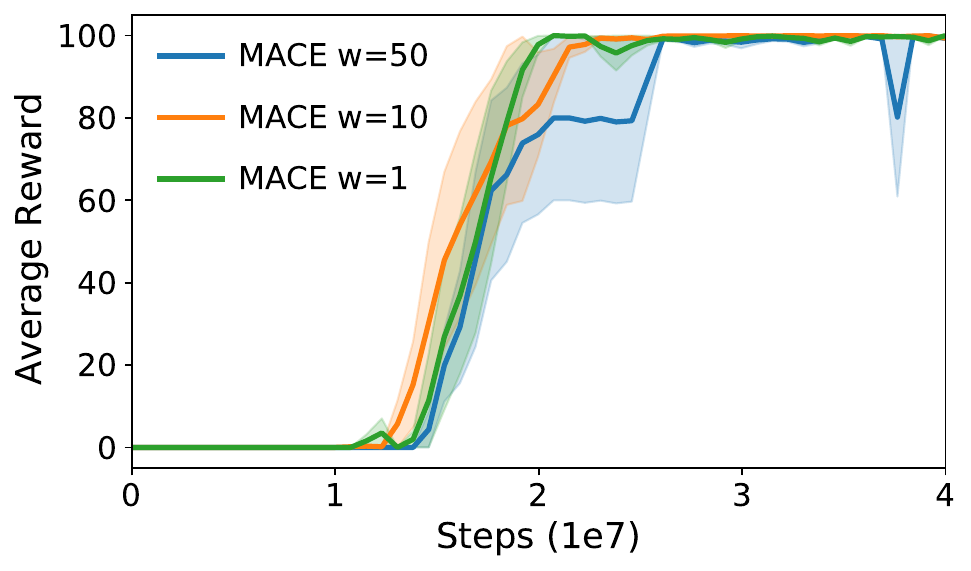}
    }
    \qquad
    \subfigure[\texttt{SecretRoom}]{
        \includegraphics[width=0.27\columnwidth]{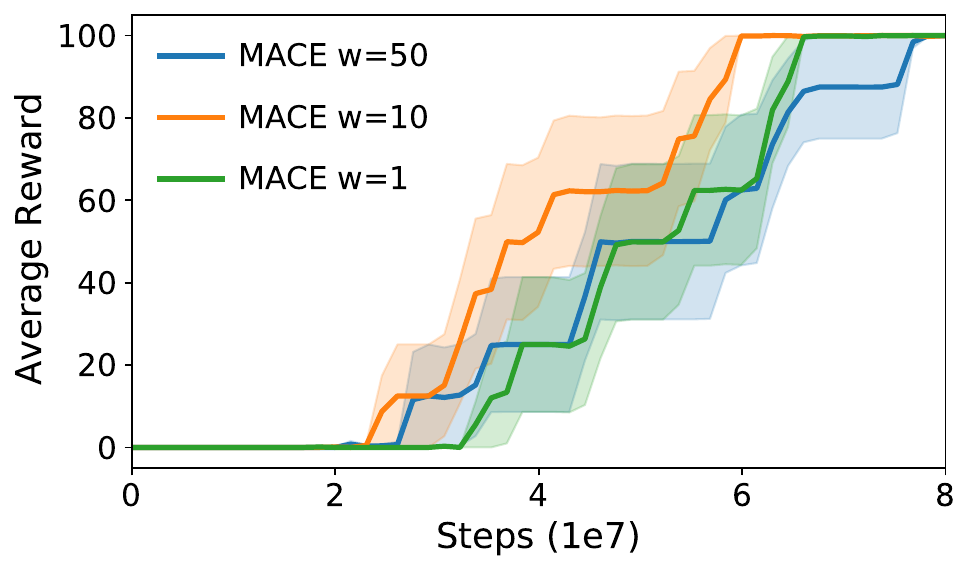}
    }
    \qquad
    \subfigure[\texttt{MultiRoom}]{
        \includegraphics[width=0.27\columnwidth]{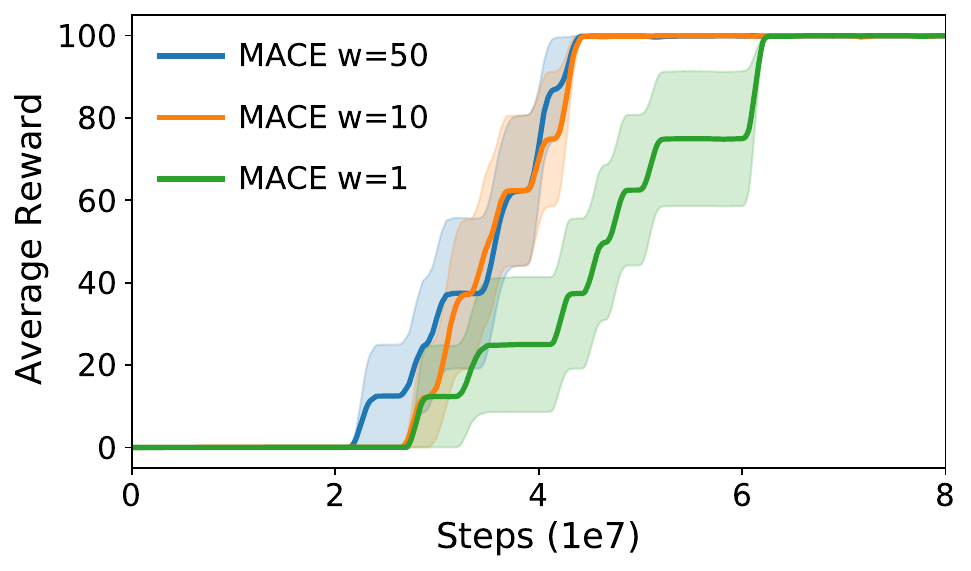}
    }
    \vskip -0.1in
    \caption{Learning curves of MACE with different $w$ on three GridWorld tasks: (a) \texttt{Pass}, (b) \texttt{SecretRoom}, and (c) \texttt{MultiRoom}. }
    \label{fig:gridworld_wsize}
    \vskip -0.1cm
\end{figure}

\cref{fig:gridworld_wsize} shows the performance of MACE with different $w$ including 1, 10, and 50, on three GridWorld tasks. We can observe that MACE with $w=10$ performs best across three tasks. MACE with $w=1$, which estimate \eqref{equ:est_pos} with on-policy samples, performs worse than MACE with $w=10$ on \texttt{SecretRoom} and \texttt{MultiRoom}. MACE with $w=50$ performs worse than MACE with $w=10$ on \texttt{Pass} and \texttt{SecretRoom}. These results show: On the one hand, using on-policy data to estimate \eqref{equ:est_pos} is unbiased, but it may cause a large variance due to the small amount of data. On the other hand, estimating \eqref{equ:est_pos} by combining off-policy data with on-policy data could increase the amount of data and reduce the variance, but it is biased. Therefore, $w$ controls the balance between bias and variance.

\subsection{MACE with Different $\lambda$}
\label{app:lambda}

\begin{figure}[htb]
    \centering
    \subfigure[\texttt{Pass}]{
        \includegraphics[width=0.27\columnwidth]{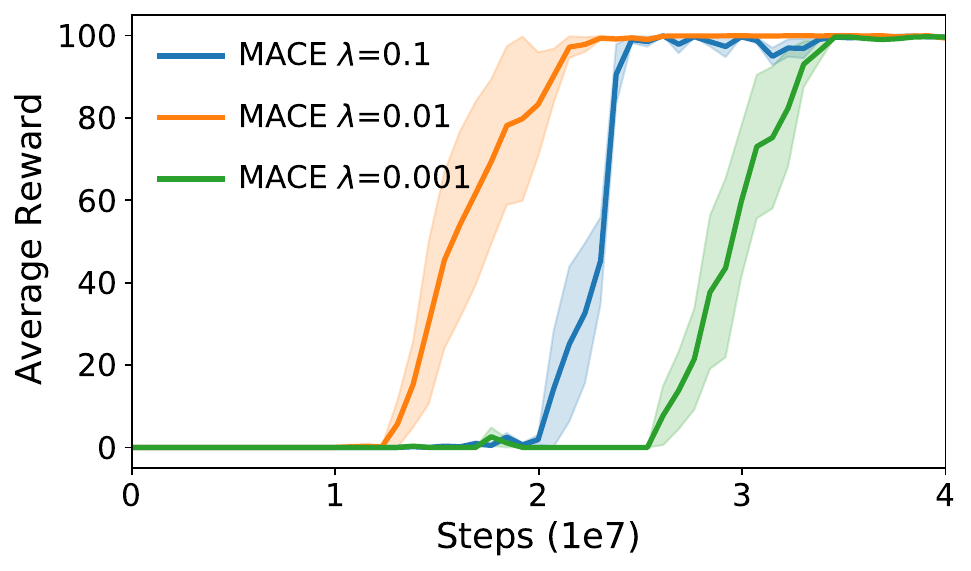}
    }
    \qquad
    \subfigure[\texttt{SecretRoom}]{
        \includegraphics[width=0.27\columnwidth]{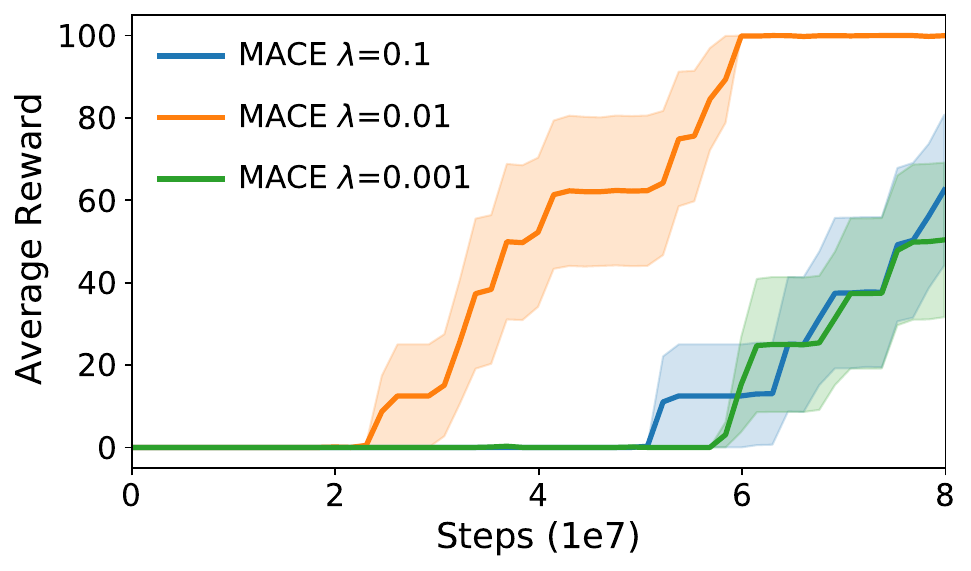}
    }
    \qquad
    \subfigure[\texttt{MultiRoom}]{
        \includegraphics[width=0.27\columnwidth]{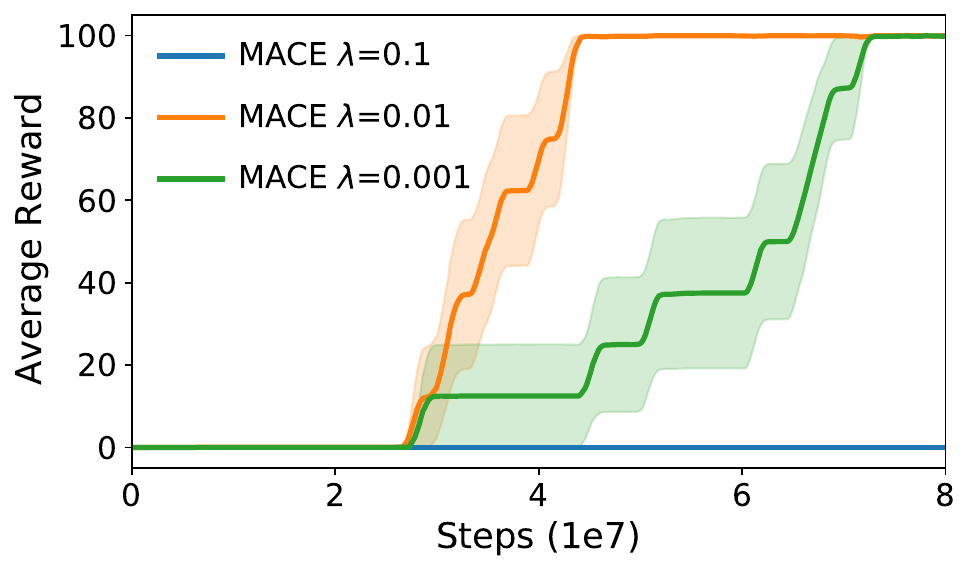}
    }
    \vskip -0.1in
    \caption{Learning curves of MACE with different $\lambda$ on three GridWorld tasks: (a) \texttt{Pass}, (b) \texttt{SecretRoom}, and (c) \texttt{MultiRoom}. }
    \label{fig:gridworld_lambda}
    \vskip -0.1cm
\end{figure}

\cref{fig:gridworld_lambda} shows the performance of MACE with different $\lambda$ including 0.1, 0.01, and 0.001, on three GridWorld tasks. MACE with $\lambda=0.01$ outperforms MACE with $\lambda=0.1$ and MACE with $\lambda=0.001$ significantly across all tasks. Therefore, $\lambda=0.01$ can maintain a good balance between the novelty-based intrinsic reward encouraging the agent to globally novel states and the hindsight-based intrinsic reward encouraging the agent to influence other agents' exploration. As shown in Table~\ref{tab:specific_hp}, in Overcooked, $\lambda$ is set to be $0.1$ across all tasks. However, in SMAC, MACE has a different $\lambda$ for \texttt{2m\_vs\_1z}, \texttt{3m} and \texttt{8m}. The main reason is that each map in SMAC may require a different strategy to win the game. Coordinated exploration is likely to play a more important role in \texttt{2m\_vs\_1z} than in \texttt{3m} and \texttt{8m}. In more detail, the winning strategy in \texttt{2m\_vs\_1z} is alternating fire, where the Marine chased by Zealot keeps running away, and the other Marine fires at Zealot. So agents in this task require highly coordinated exploration. In contrast, exploration of agents are more independent in \texttt{3m} and \texttt{8m}, where the agent's behavior does not affect other agents firing at enemies. We can also observe this difference from \cref{fig:sc_result}: IPPO with the local novelty reaches around 37.5\% winning rate (3 out of 8 seeds reach 100\% winning rate) in \texttt{3m}, but cannot learn at all in \texttt{2m\_vs\_1z}, verifying that exploration is more independent in \texttt{3m} than in \texttt{2m\_vs\_1z}. Regarding the poor performance of IPPO+r\_loc in \texttt{8m}, we speculate that the reason may be the increasing gap between the local novelty and the global novelty with the increasing number of agents. Besides, the interaction between agents becomes more complicated in \texttt{8m}, making coordination more important and requiring a higher $\lambda$ than \texttt{3m}.

\subsection{Comparison with More Baselines}
\label{app:ctde}

\begin{figure}[htb]
    \centering
    \subfigure[\texttt{Pass}]{
        \includegraphics[width=0.27\columnwidth]{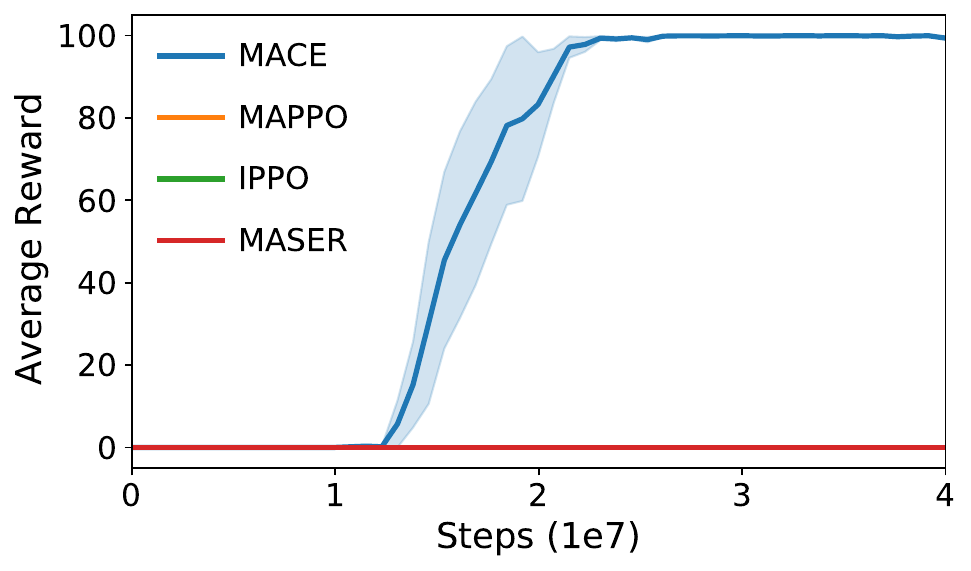}
    }
    \qquad
    \subfigure[\texttt{SecretRoom}]{
        \includegraphics[width=0.27\columnwidth]{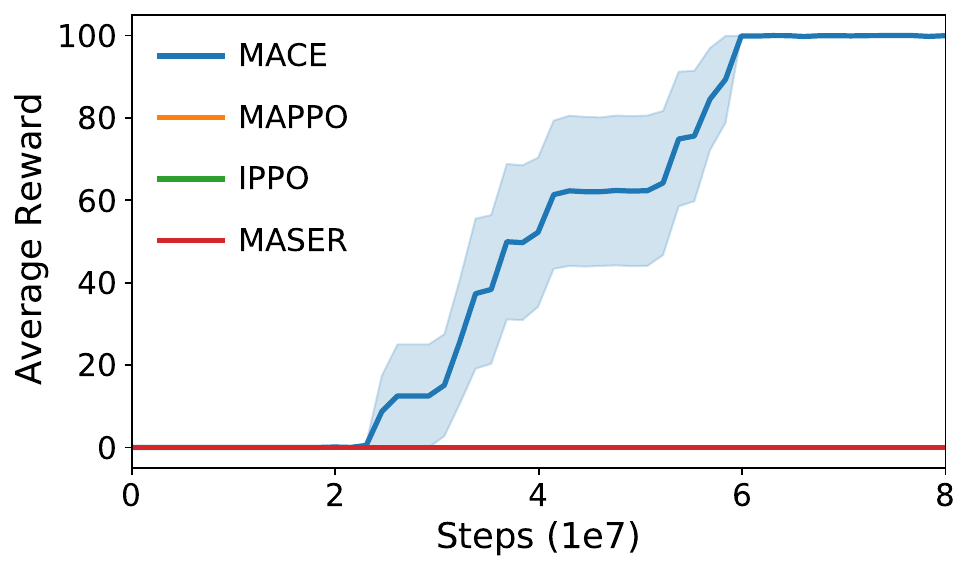}
    }
    \qquad
    \subfigure[\texttt{MultiRoom}]{
        \includegraphics[width=0.27\columnwidth]{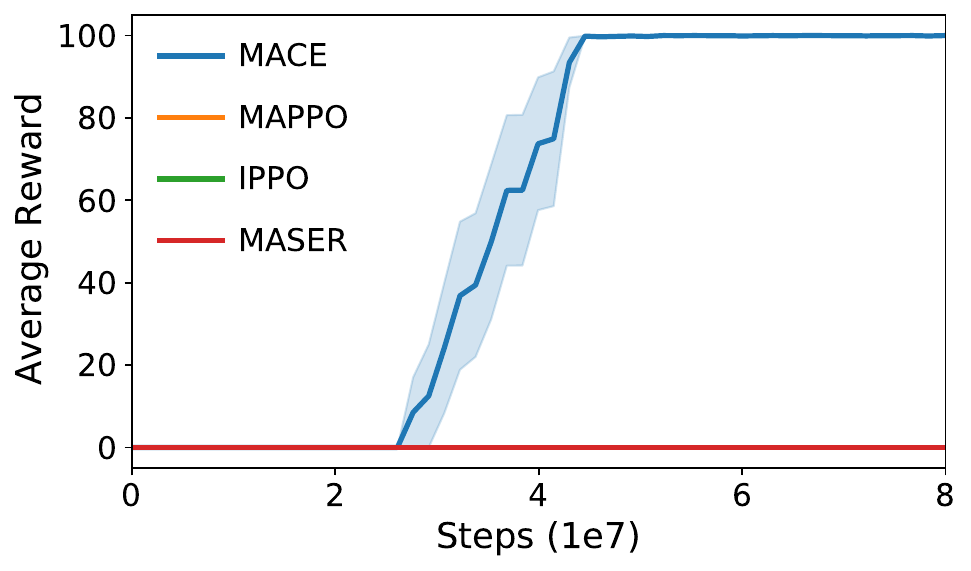}
    }
    \vskip -0.1in
    \caption{Learning curves of MACE compared with IPPO, MAPPO, and MASER on three GridWorld tasks: (a) \texttt{Pass}, (b) \texttt{SecretRoom}, and (c) \texttt{MultiRoom}. }
    \label{fig:gridworld_vc}
    \vskip -0.1cm
\end{figure}

We compared MACE with some additional baselines on three GridWorld tasks. These baselines include: a) IPPO: the basic decentralized RL algorithm we used in this paper; b) MAPPO \cite{yu2021surprising}: a popular simple-yet-effective CTDE method; c) MASER \cite{jeon2022maser}: a recent SOTA centralized multi-agent exploration method. The results shown in \cref{fig:gridworld_vc} indicate that none of IPPO, MAPPO, and MASER succeeds on GridWorld tasks. IPPO and MAPPO fail as a result of the sparse reward signal and the absence of motivation to explore. To our surprise, MASER also shows ineffective. We speculate that its ineffectiveness could be attributed to the method itself and its default hyperparameters, as we used here following the provided code\footnote{\url{https://github.com/Jiwonjeon9603/MASER}}, which appears to be tailored specifically for SMAC tasks. CMAE \cite{liu2021cooperative} is another contemporary centralized multi-agent exploration method. Figure 1 in CMAE paper presents its performance on \texttt{Pass} and \texttt{SecretRoom}. CMAE proves better sample efficiency than MACE on these two tasks due to its off-policy learning, tabular Q-function, and access to the global state.

We also carried out an experiment to compare MACE with these baselines on the \texttt{2m\_vs\_1z} with sparse reward. The result averaged over 8 seeds, is presented in \cref{fig:2m_vs_1z_cmp}, which shows that MACE achieves a higher win rate than MASER, indicating that our proposed intrinsic rewards are more successful in encouraging cooperative exploration on \texttt{2m\_vs\_1z}. MAPPO and IPPO also fail to learn a valid cooperative policy on this map due to sparse-reward.

\subsection{Dense Reward}
\label{app:dense}

\begin{figure}[htb]
    \centering
    \subfigure[\texttt{2m\_vs\_1z}]{
        \includegraphics[width=0.27\columnwidth]{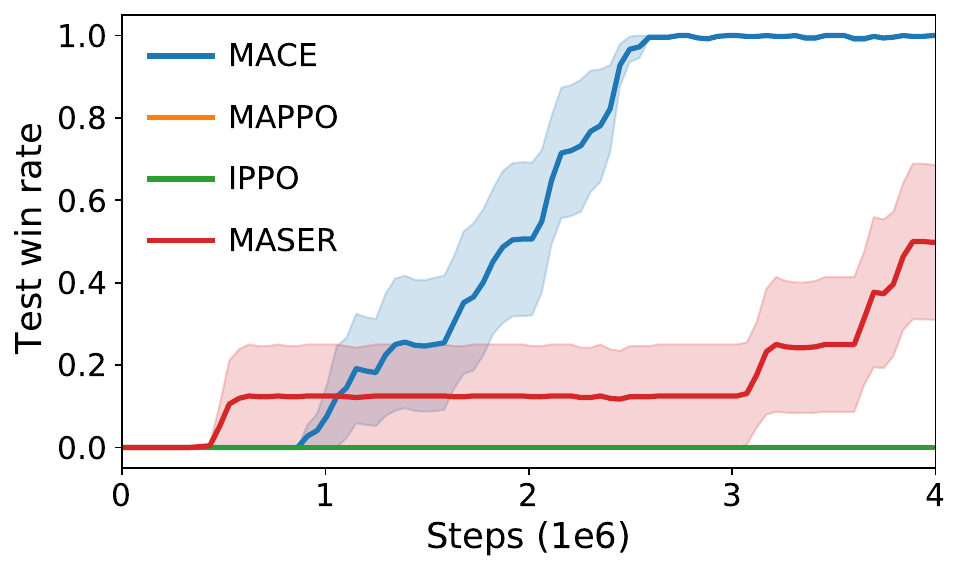}
        \label{fig:2m_vs_1z_cmp}
    }
    \qquad
    \subfigure[\texttt{3s\_vs\_5z}]{
        \includegraphics[width=0.27\columnwidth]{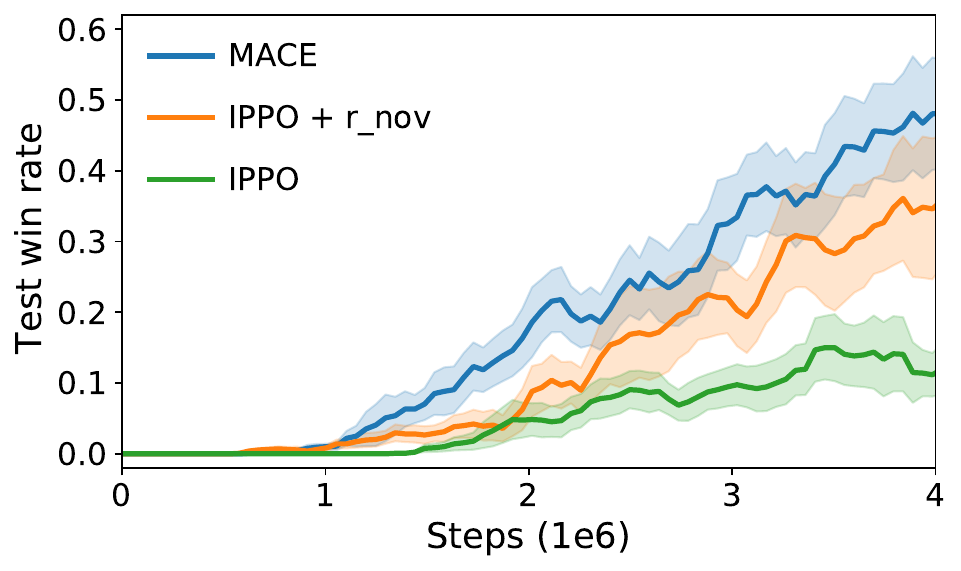}
        \label{fig:3s_vs_5z} 
    }
    \qquad
    \subfigure[\texttt{MultiRoom}]{
        \includegraphics[width=0.27\columnwidth]{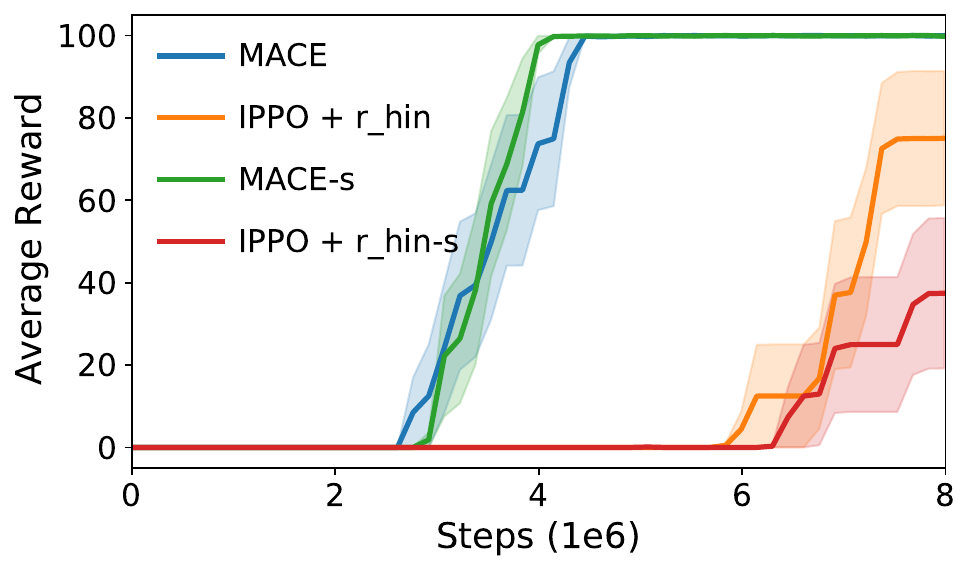}
        \label{fig:scale}
    }
    \caption{(a) Learning curves of MACE compared with MAPPO, MASER, and IPPO on \texttt{2m\_vs\_1z} with sparse reward. (b) Learning curves of MACE compared with IPPO+r\_nov, and IPPO on \texttt{3s\_vs\_5z} with dense reward. (c) Learning curves of MACE compared with MACE-s, IPPO+r\_hin, and IPPO+r\_hin-s on \texttt{MultiRoom}.
}
\end{figure}

To verify whether MACE can work well on hard tasks with dense reward, we carried out an experiment to compare MACE with IPPO on a hard SMAC map, \texttt{3s\_vs\_5z}, with normal reward. As the normal reward is scaled to a maximum of 20 (default in SMAC), we introduce a new hyperparameter $\beta$ to scale our intrinsic rewards used in \cref{equ:final_r}: 
\begin{equation}
    r_\mathrm{s}^i=r_\mathrm{ext}+\beta(r_\mathrm{nov}^i+\lambda \sum_{j\neq i}r_\mathrm{hin}^{i\rightarrow j})
\end{equation}
We set $\beta$ to 0.1 and $\lambda$ to 0.01, while keeping other hyperparameters the same as those used in 3m. The result, averaged over 8 seeds, is shown in \cref{fig:3s_vs_5z}. MACE outperforms IPPO significantly, proving that MACE also works well on hard tasks with dense reward. Furthermore, the win rate of MACE is slightly higher than that of our ablation IPPO + r\_nov, indicating the effectiveness of using hindsight-based intrinsic rewards to facilitate coordinated exploration in normal dense reward tasks.

\section{Scalability}
\label{app:scale}

According to the definition of the hindsight-based intrinsic reward \eqref{equ:final_rij} and the estimation of posterior distribution $p(a_t^i|o_t^i,z_t^j)$ described in \cref{app:est_pos}, each agent requires $N-1$ repeated modules (tables in GridWorld, and $f_p$s in Overcooked and SMAC) if there are $N$ agents in the environment. Each module describes a conditional distribution of the agent's action on one other agent's accumulated novelty and its own observation. Therefore, the size of each agent's model increases linearly with the number of agents in the environment. To guarantee the scalability of MACE when extending to more agents, we propose a scalable hindsight-based intrinsic reward $r^i_\mathrm{hin-s}$. Instead of weighted mutual information between agent $i$'s action $a_t^i$ and agent $j$'s accumulated novelty $z_t^j$ given agent $i$'s observation $o_t^i$, we use weighted mutual information between $a_t^i$ and $z_t^{-i}=\sum_{j\neq i}z_t^j$, the summation of all other agents' $z_t^j$, given $o_t^i$:
\begin{equation}
\label{equ:wmi-s}
    \omega I\left(A_t^i;Z^{-i}_t | o_t^i\right) =
    \mathbb{E}_{a_t^i,z_t^{-i}| o_t^i}\left[z_t^{-i}\log\frac{p(a_t^i,z_t^{-i}| o_t^i)}{p(a_t^i| o_t^i)p(z_t^{-i} | o_t^i)}\right].
\end{equation}
Then we can define the intrinsic reward $r^i_\mathrm{wmi}(o_t^i)=\omega I\left(A_t^i;Z^{-i}_t | o_t^i\right)$, decompose it, and get the scalable hindsight-based intrinsic reward:
\begin{equation}
\label{equ:s_ri}
    r^i_\mathrm{hin-s}(o_t^i,a_t^i,\{z_t^j\}_{j\neq i})=z_t^{-i}\log\frac{p(a_t^i|o_t^i,z_t^{-i})}{\pi^i(a_t^i|o_t^i)}.
\end{equation}
Compared to \eqref{equ:final_rij}, \eqref{equ:s_ri} reduces computation overhead, because each agent requires only one module to represent the conditional distribution of the agent's action on the summation of all other agent's accumulated novelty and its own observation.

We test the scalable hindsight-based intrinsic reward on \texttt{MultiRoom}, where $z_t^{-i}$ is the summation of the other two agents' accumulated novelty. \cref{fig:scale} shows the performance of MACE, IPPO+r\_hin, MACE-s, and IPPO+r\_hin-s, in which MACE-s and IPPO+r\_hin-s replace $r_\mathrm{hin}^i$ in MACE and IPPO+r\_hin with $r_\mathrm{hin-s}^i$ respectively. The learning curves of MACE and MACE-s are close, verifying the effectiveness of the scalable hindsight-based intrinsic reward. IPPO+r\_hin performs better than IPPO+r\_hin-s, suggesting that $r_\mathrm{hin}^i$ may be more accurate and effective than $r_\mathrm{hin-s}^i$ alone.

\end{document}